\documentclass[11pt]{article}
\usepackage{latexsym}

\def\be{\begin{equation}}
\def\ee{\end{equation}}
\def\bea{\begin{eqnarray}}
\def\eea{\end{eqnarray}}
\def\pa{\partial}

\def\fn{\footnote}
\def\vc{V^{\frac{2}{3}}}

\def\case#1/#2{\textstyle\frac{#1}{#2}}

\begin{document}
\begin{titlepage}

\vspace{.7in}

\begin{center}
\Large\bf {SCALE-INVARIANT GRAVITY: GEOMETRODYNAMICS}\normalfont
\\
\vspace{.4in} \normalsize \large{Edward Anderson$^1$, Julian
Barbour$^2$, Brendan  Foster$^3$ and Niall \'{O} Murchadha$^4$}
\\
\normalsize \vspace{.4in}
$^1${\em  Astronomy Unit, School of Mathematical Sciences, \\
Queen Mary, Mile End Road, London E1 4NS, U.K.

$^2$ College Farm, South Newington, Banbury, Oxon, OX15 4JG, UK

$^3$ Physics Department, Univ. of Maryland,
College Park, Maryland, USA

$^4$ Physics Department, University College, Cork, Ireland}

\end{center}
\vspace{.4in} 

\begin{abstract}

We present a scale-invariant theory, \textit{conformal gravity}, 
which closely resembles the geometrodynamical formulation of general 
relativity (GR).
While previous attempts to create scale-invariant theories of gravity have 
been based on
Weyl's idea of a compensating field, our direct approach dispenses with this 
and is built
by extension of the method of best matching w.r.t scaling developed in the 
parallel particle dynamics paper by one of the authors.  
In spatially-compact GR, there is an infinity of degrees of freedom that 
describe the shape
of 3-space which interact with a single volume degree of freedom.  In 
conformal gravity,
the shape degrees of freedom remain, but the volume is no longer a dynamical 
variable.  Further theories and formulations related to GR and conformal gravity are  
presented.

Conformal gravity is successfully coupled to scalars and the gauge fields of 
nature.
It should describe the solar system observations as well as GR does, but its 
cosmology
and quantization will be completely different.

\end{abstract}

\vspace{.3in} 

Electronic addresses:

$^1$ {eda@maths.qmw.ac.uk}, {$^2$ julian@platonia.com,} {$^3$
bzf@glue.umd.edu,} {$^4$ n.omurchadha@ucc.ie}

\end{titlepage}

\section{Inroduction}

During the years that he created special and general relativity,
Einstein had several goals \cite{spacetime}.  The first, realized
in special relativity, was to reconcile Maxwell's wave theory of
light with universal validity of the restricted relativity
principle (RRP). In contrast to Lorentz, who explicitly sought a
constructive theory \cite{constructive} to explain the Michelson--Morley experiment
and the RRP, Einstein was convinced that the quantum effects discovered by Planck 
invalidated such an approach \cite{spacetime}, p. 45. He ``despaired of the possibility 
of discovering the true laws by means of constructive efforts'' 
\cite{spacetime}, p. 53, and instead adopted the RRP as an axiomatic principle.  
Einstein's further goals were the implementation of Mach's principle and 
the construction of a field theory of gravitation analogous to Maxwellian
electromagnetism. Encouraged by his treatment of the RRP as a
principle to be adopted rather than a result to be derived,
Einstein generalized it to the general relativity principle (GRP),
according to which the laws of nature must take an identical form
in all frames of reference. The GRP was eventually
implemented as the four-dimensional general covariance of a
pseudo-Riemannian dynamical spacetime.

In making spacetime the arena of dynamics, Einstein broke
radically with the historical development of dynamics, in which
the configuration space and phase space had come to play ever more
dominant roles. In fact, both of these played decisive roles in
the discovery of quantum mechanics, especially the symplectic
invariance of Hamiltonian dynamics on phase space.  Since then,
Hamiltonian dynamics has also played a vital role in the emergence
of modern gauge theory \cite{Dirac}. In fact, spacetime and the canonical
dynamical approach have now coexisted for almost a century, often
creatively but also not without tension.

This tension became especially acute when Dirac and Arnowitt,
Deser and Misner (ADM) \cite{Dirac, ADM} reformulated the Einstein
field equations as a constrained Hamiltonian dynamical system
describing the evolution of Riemannian 3-metrics $g_{ij}$:
\be
{\mbox{\sffamily H\normalfont}} = \int \textrm{d}^3x (N{\cal H} +
\xi^i{\cal H}_{i}) \label{ADM} \ee \be {\cal H} \equiv
G_{ijkl}p^{ij}p^{kl} - \sqrt{g}R = 0 , \label{ham}
\ee
\be
{\cal
H}_i \equiv -2\nabla_jp_i{}^j = 0 , \label{mom}
\ee
where a
divergence term has been omitted from (\ref{ADM}). The 3-metric
$g_{ij}$\fn{In this paper, we use lower case Latin letters for
spatial indices and upper case Latin letters for internal indices.
We use round brackets for symmetrization and square brackets for
antisymmetrization; indices to be excluded from
(anti)symmetrization are set between vertical lines.} has
determinant $g$ and conjugate momentum $p^{ij}$, $N$ is the lapse,
$\xi^i$ is the shift, $R$ is the 3-dimensional Ricci scalar,
$\nabla_j$ is the 3-dimensional covariant derivative, $G_{ijkl} =
\frac{1}{\sqrt{g}}(g_{i(k}g_{l)j} - \frac{1}{2}g_{ij}g_{kl})$ is
the DeWitt supermetric \cite{DeWitt}, and ${\cal H}$, ${\cal H}_i$
are the algebraic scalar Hamiltonian constraint and differential
vector momentum constraint respectively. Dirac was so impressed by
the simplicity of the Hamiltonian formulation that he questioned
the status of spacetime \cite{DiracPRC}, remarking `I am inclined
to believe \dots that four-dimensional symmetry is not a
fundamental property of the physical world.' Wheeler too was
struck by this development and coined the words
\textit{geometrodynamics} for the Einsteinian evolution of
3-dimensional Riemannian geometries (\textit{3-geometries})
embedded in spacetime and \textit{superspace}, the configuration
space formed by all 3-geometries on a given 3-manifold $\cal M$
\cite{Battelle}.

Mathematically, superspace is obtained from Riem, the
space of all (suitably continuous) Riemannian 3-metrics $g_{ij}$
defined on $\cal M$ (taken here -- as an important physical
assumption --  to be compact and without
boundary) by quotienting with respect to 3-dimensional
diffeomorphisms on $\cal M$: \be \mbox{\{Superspace\}} = \frac
{\mbox {\{Riem\}} } {\mbox {\{3-Diffeomorphisms\}} }.
\label{Superspace} \ee Superspace is the analogue of the relative
configuration space discussed in the particle-dynamics paper \cite{PD}.  
\fn{This paper is henceforth referred to as PD.  
References to it are identified by PD followed by
the relevant section or equation number.}

The ADM configuration space is the extension of Riem
to $\mbox{Riem} \times \Xi \times P$, where $\Xi$ is the space of the $\xi^i$ 
and $N \in P$, the space of (suitably differentiable) positive functions.
However, since $N$ and $\xi^i$ have no conjugate momenta, the true
gravitational degrees of freedom of GR are contained in
Riem. These degrees of freedom are furthermore
subjected to the Hamiltonian and momentum constraints (\ref{ham})
and (\ref{mom}). If, in the thin-sandwich problem (Sec. PD.4) for
geometrodynamics (Sec. 2), one could solve the momentum constraint
in terms of the Lagrangian variables for $\xi^i$, then the theory,
now formally defined on Riem, would actually be defined on
superspace because the physical degrees of freedom have been
reduced to three. However, on account of the still remaining
Hamiltonian constraint (\ref{ham}), superspace must still contain
one redundant degree of freedom per space point.

The hitherto unresolved status of this one remaining redundancy,
which can only be eliminated at the price of breaking the
spacetime covariance of GR, is probably the reason why neither
Dirac nor Wheeler, despite being struck by the Hamiltonian
structure of GR, made any serious subsequent attempt to do without
the notion of spacetime. In particular, Wheeler formulated the
idea of embeddability, i.e, that Riemmanian 3-geometries always
evolve in such a way that they can be embedded in a
four-dimensional pseudo-Riemannian spacetime \cite{Battelle}. This
idea led Hojman, Kucha$\check{\textrm r}$ and Teitelboim
\cite{HKT} to a new derivation of general relativity as a
constrained Hamiltonian system. However, this was not a purely
dynamical derivation, since the embeddability condition played a
crucial role.

In the recent \cite{BOF}, called henceforth RWR from its title
`Relativity Without Relativity', and \cite{AB}, we showed that GR,
the universal light cone of special relativity, and the gauge
principle could all be derived in a unified manner using
principles that in no way presuppose spacetime. Our so-called
3-space approach developed out of an earlier attempt \cite{BB} to
implement Mach's idea of a relational dynamics directly in a
constructive dynamical theory, in contrast to Einstein's indirect
approach through generalization of the RRP to the GRP. For the
direct approach, superspace is the natural relational arena for
geometrodynamics. A detailed motivation is given in \cite{BOF} and
PD.  The aim of this paper is to extend the techniques of \cite{BOF} 
to include scaling invariance and thereby push the idea of relational 
dynamics to its logical
conclusion. This has already been achieved in PD for the
Newtonian context of point particles that interact in Euclidean
space. The present paper extends the scaling techniques developed
in PD to geometrodynamics.  This adds to the aforementioned
tension, since it provides an example of a theory of evolving
3-geometries that is not generally covariant but couples
satisfactorily to the accepted bosonic fields of nature and should
pass the solar-system and binary-pulsar tests of GR.

As explained further in PD, the aim of the 3-space approach is, first,
to identify the configuration space Q$_{0}$ of the true physical
degrees of freedom and, second, to formulate the dynamics in such
a way that specification of an initial point $q_{0}$ in Q$_{0}$
and an initial \textit{direction} at $q_{0}$ suffices to determine
a unique dynamical curve in Q$_{0}$. This relational geodesic
principle is implemented by \textit{best matching}, which is
explained in detail in \cite{BOF} and PD, using an action that
is \textit{homogeneous of degree one in the velocities}.

Both properties of the action -- the best matching and the
homogeneity -- lead to constraints.  An important tool of the
3-space approach is Dirac's generalized Hamiltonian dynamics
\cite{Dirac}, used to ensure that all the constraints are propagated by
the Euler--Lagrange equations. The precise form of the best
matching, which always leads to constraints linear and homogeneous
in the canonical momenta, is either clear in advance, being
determined by the symmetry used in the quotienting to obtain the
physical configuration space Q$_{0}$ [a good example is the
quotienting by 3-diffeomorphisms, which leads to superspace
(\ref{Superspace}) and the momentum constraint (\ref{mom})] or is
dictated by the emergence of a constraint when the Dirac
consistency procedure is used (this is how gauge theory arises of
necessity and is then encoded in the action in RWR \cite{BOF,
AB}).

For the homogeneity, there is an important freedom in the manner
in which it is implemented. In PD, the Lagrangian is a simple
square root of an expression quadratic in the velocities. In
\cite{BOF,AB} and the present paper, a \textit{local square root}
is used. This means that the Lagrangian is an integral over the
manifold $\cal M$ of the square root of a quadratic expression
calculated at each point of $\cal M$ before the integration is
performed. The local square root leads to infinitely many
quadratic constraints, one per  point of ${\cal M}$. This is in
marked contrast to the single `global' quadratic constraint that
follows from Jacobi-type actions in particle mechanics.

In RWR \cite{BOF}, the local square root is in the
Baierlein--Sharp--Wheeler (BSW) form \cite{BSW} for the action for
GR, \be S_{\mbox{\scriptsize BSW\normalsize}} = \int
\textrm{d}\lambda\int \textrm{d}^3x \sqrt{g}\sqrt{R}\sqrt{T},
\label{BSWac} \ee where the kinetic term \be T =
\frac{1}{\sqrt{g}}G^{abcd}\frac{ \textrm{d} g_{ab}}{\textrm{d}
\lambda}\frac{\textrm{d} g_{cd}}{\textrm{d} \lambda}
\label{KE}\ee is constructed by the best-matching correction of
the velocities to allow for the action of 3-diffeomorphisms (which
generalize simultaneously the translations and rotations
considered for particles in Euclidean space in PD)\fn{As in
PD, we shall use the formalism of corrected coordinates and
corrected velocities to treat conformal transformations. For the
3-diffeomorphisms, we shall use only corrected velocities,
denoting them, as in (\ref{CorrVel}), by the `straight' d. Note
also that in this paper we use the widely employed ADM sign
conventions and notation, which results in minor differences
compared with the expressions in RWR \cite{BOF}.} \be
\frac{\textrm{d} g_{ab}}{\textrm{d}\lambda} \equiv \frac{\pa
g_{ab}}{\pa\lambda} - (\xi^c\pa_cg_{ab} + g_{ac}\pa_b\xi^c +
g_{bc}\pa_a\xi^c) = \frac{\pa g_{ab}}{\pa\lambda} 
- \pounds_{\xi}g_{ab}, \label{CorrVel} \ee and the inverse DeWitt
supermetric  $G^{abcd} = \sqrt{g}(g^{ac}g^{bd} - g^{ab}g^{cd})$.  
Here, $\pounds_{\xi}$ is the Lie derivative w.r.t $\xi_i$.  
We often abbreviate $\frac{\pa}{\pa\lambda}$ by a dot.  

The 3-space approach works in RWR \cite{BOF} as follows. One
starts with an action of the same form as (\ref{BSWac}) but with
the 3-scalar curvature $R$ replaced by an arbitrary scalar
concomitant of the 3-metric $g_{ij}$ and the general inverse supermetric
$G^{ijkl}_{\mbox{\scriptsize\textrm{W}\normalsize}}
= \sqrt{g}(g^{ik}g^{jl} - Wg^{ij}g^{kl})$
for \textit{W} constant ($W \neq \frac{1}{3}$ is required for invertibility).
At each point of $\cal M$,
a scalar algebraic constraint and a vector differential constraint
must be satisfied. The scalar constraint arises purely from the
local square-root form of any BSW-type Lagrangian, is quadratic in
the canonical momenta and has the general form of the quadratic
ADM Hamiltonian constraint ${\cal H}$ (\ref{ham}). The vector
constraint arises from variation w.r.t the auxiliary variable
$\xi^{i}$ and is identical to the ADM momentum constraint ${\cal
H}_i$ (\ref{mom}). This universal form arises because best
matching entails replacement of the `bare' velocity
$\partial{g_{ij}}/\partial\lambda$ by the corrected velocity
$\textrm{d}g_{ij}/\textrm{d}\lambda$ (\ref{CorrVel}), the form of
which is uniquely determined by the diffeomorphism symmetry. One
then checks whether the modified Euler--Lagrange equations
propagate the constraints. In general, they do not, and to arrive
at a consistent theory one must either introduce new constraints
as outlined by Dirac \cite{Dirac} or drastically limit the
modifications of the BSW action. However, new constraints will
rapidly exhaust the degrees of freedom, which have already been
reduced to two per space point by the existing constraints. This
may be called Dirac's method by exhaustion, and it is very
powerful. In RWR \cite{BOF} it was shown to enforce $W = 1$ and to
limit the possible modifications of $R$ to $sR + \Lambda$, where
$\Lambda$ is a cosmological constant and $s=0,1,-1$, corresponding
to so-called strong gravity \cite{stronggravity} (for which $W\neq
1$ is allowed \cite{Sanderson}), and Lorentzian and Euclidean GR,
respectively. This is the `relativity without relativity' result.
This and other modifications of the action were considered in
\cite{BOF,BO,Sanderson}. Consistency also exhaustively forces
classical bosonic theories to have the form of the currently known
bosonic gauge fields and to respect a universal light cone
\cite{BOF,AB,BOF2}.

In this paper, we construct a new scale-invariant theory of
gravity by employing best matching not only w.r.t
3-diffeomorphisms but also w.r.t 3-dimensional conformal
transformations: \be g_{ij}\longrightarrow \omega^{4}g_{ij},
\label{ConfTrans} \ee where the scalar $\omega$ is an arbitrary
smooth positive function of the label $\lambda$ and of the
position on ${\cal M}$. The transformation (\ref{ConfTrans}) is
the infinite dimensional `localization' of the first member of the
scaling transformation (Eq. PD.23). (The fourth power of $\omega$ is
traditional and is used for computational convenience.) The new
theory, which we call \textit{conformal gravity}, is a consistent
best-matching generalization of the BSW action invariant under
(\ref{ConfTrans}). It was proposed in the brief communication
\cite{BO}, which, however, treated only pure gravity and was
written before the development of the special variational
technique with free end points described in PD. The present
paper uses the new method and develops three aspects: the Hamiltonian formulation, the coupling to
the known bosonic matter fields and the physical and mathematical implications of conformal gravity.

In Hamiltonian GR, the six degrees of freedom per manifold point
present in Riem are reduced to two by the
differential vector constraint (\ref{mom}) and the algebraic
scalar Hamiltonian constraint (\ref{ham}). These constraints arise
from variation w.r.t to the lapse $N$ and shift $\xi^{i}$. The
lapse and the shift remain completely free gauge variables and can
be chosen at will in the course of evolution. They are velocities
in unphysical gauge directions.

In conformal gravity, we retain the basic form of the BSW action
with local square root but replace $g_{ij}$ by the corrected
coordinates
\be
    \bar g_{ij}=\phi^4g_{ij}
\label{CoCo} \ee and add the best matching w.r.t the conformal
factor $\phi$ in (\ref{CoCo}). This leads to a second algebraic
scalar constraint that holds at each space point:\be
\textrm{tr}p^{ij}\equiv p^{ij}g_{ij}\equiv p=0, \label{ConCon}\ee
which, unlike the quadratic Hamiltonian constraint, is
\textit{linear} in the canonical momenta. The trace $p$ in
conformal gravity is analogous to the dilatational momentum defined by (Eq. PD.3).
However, since (\ref{ConfTrans}) is a local transformation, in
contrast to the global transformation (Eq. PD.23), we have one
conformal constraint at each space point as with the Hamiltonian
and momentum constraints. Just as the vanishing dilatational momentum in particle
dynamics conserves the moment of inertia $I$ of an island
universe, here the vanishing trace conserves the volume of a
spatially compact universe.

Since conformal gravity augments the ADM Hamiltonian (\ref{ham})
and momentum (\ref{mom}) constraints by the conformal constraint
(\ref{ConCon}), a simple count suggests that the new theory should
have only a single true degree of freedom per space point, in
contrast to the two present in Hamiltonian GR. However, this is
not the case. We show in Sec. 3 that the free-end-point variation
w.r.t the conformal factor $\phi$ leads not only to the
constraint (\ref{ConCon}) but also to a further condition that
fixes the analogue in conformal gravity of the lapse $N$ in GR.
Whereas in GR $N$ and $\xi^i$ are freely specifiable gauge
velocities, in conformal gravity $N$ is fixed, and its role as
gauge variable is taken over by $\phi$. Since $\xi^i$ plays the
same role in both theories, conformal gravity, like GR, has two
degrees of freedom per space point. However, in contrast to GR,
they are unambiguously identified as the two conformal shape
degrees of freedom of the 3-metric $g_{ij}$. Conceptually, this is
a pleasing result, but it has a far-reaching consequence --
conformal gravity cannot be cast into the form of a
four-dimensionally generally covariant spacetime theory.  Because
the lapse is fixed, absolute simultaneity and a preferred frame of
reference are introduced.

The reader may feel that this is too high a price to pay for a
scale-invariant theory.  Of course, experiment will have the final
word. However, one of the aims of this paper is to show that
conformal invariance already has the potential to undermine the
spacetime covariance of GR. In order to demonstrate this, in Sec.
2 we explain York's powerful conformal method \cite{York71,
York72, York73, York74} for finding initial data that satisfy the
constraints (\ref{ham}) and (\ref{mom}). This serves three
purposes: it introduces invaluable concepts and techniques, it
establishes an intimate connection between GR and conformal
gravity, and it facilitates the testing of conformal gravity as a
putative description of nature, since the two theories are most
easily compared using York's techniques.

York's work was stimulated in part by Wheeler's desire to find the
true degrees of freedom of GR and with them the physical
configuration space of geometrodynamics. For superspace still
contains redundancy since it possesses three degrees of freedom
per space point whereas GR has only two. In answer to Wheeler's
question ``What is two thirds of superspace?'', York responded
that there is only one simple and natural answer. He noted that
one can decompose an arbitrary 3-metric $g_{ij}$ into its
determinant $g$ and its scale-free part
\be
\hat{g}_{ij} \equiv g^{-{1\over 3}}g_{ij},
\label{Decomp}
\ee
which is
invariant under the conformal transformation (\ref{ConfTrans})
$\cite{York73}$. York argued that $g$ should be regarded as an
unphysical gauge degree of freedom, the elimination of which would
remove the final redundancy from GR. Though he continued to work
in spacetime, this led him to parametrize the initial data for GR in
\textit{conformal superspace} (CS) \cite{York74}, which is
obtained by quotienting Riem by both
3-diffeomorphisms and conformal transformations
(\ref{ConfTrans})\be \textrm{CS} \equiv \{\mbox{\scriptsize
Conformal Superspace\normalsize}\} =
    \frac{\{\mbox{\scriptsize Riem\normalsize }\}}
{\{\mbox{\scriptsize
3-Diffeomorphisms\normalsize}\}\{\mbox{\scriptsize Conformal
    Transformations\normalsize}\}}.
\ee Conformal superspace has been studied more recently by Fischer
and Moncrief \cite{FischMon}.

As summarized in Sec. 2, York was able to represent GR (strictly a
large subset of its solutions for the spatially compact case)
as a dynamical scheme in which the infinitely many
local shape degrees of freedom represented by $\hat{g}_{ij}$
interact with each other and with one single global variable,
which is the total volume $V$ of 3-space: \be V=\int g^{1\over
2}\textrm{d}^{3}x. \label{VolDef} \ee

Because of this extra variable, we argue that spatially compact GR does not 
have
variables in CS but rather in the marginally larger space obtained
by adjoining the volume $V$ to CS, which we call \textit{Conformal
Superspace + Volume} and abbreviate by CS+V.  It is formally
obtained from Riem by quotienting by
3-diffeomorphisms and by \textit{volume-preserving} conformal
transformations:\be \textrm{CS+V} \equiv \{\mbox{\scriptsize
Conformal Superspace + Volume\normalsize}\} =
    \frac{\{\mbox{\scriptsize Riem\normalsize }\}}
{\{\mbox{\scriptsize 3-Diffeomorphisms\normalsize}\}
\{\mbox{\scriptsize Volume-Preserving Conformal
Transformations\normalsize}\}}. \ee

The introduction of CS enables us to formulate conformal gravity
in Sec. 3 as \textit{a best-matching geodesic theory on conformal
superspace.} It determines unique curves in CS given an initial
point and an initial direction in CS.

In Sec. 4, using the Hamiltonian formulation, we show  
that spatially compact GR in York's representation
corresponds to a closely
analogous best-matching theory in CS+V: given an initial point and
an initial direction in CS+V, a unique dynamical curve is
determined. In both theories, there is a unique definition of
simultaneity. The only difference between them is that in
conformal gravity the 3-volume (\ref{VolDef}) is no longer a
dynamical variable but a conserved quantity. Although the
equations of the best-matching interpretation of GR in CS+V are
identical to York's equations, there is a difference.
The important CMC condition (Sec. 2) no longer corresponds
to a gauge fixing, as hitherto assumed, but to a physical
condition as fundamental as the ADM momentum constraint. This will
have implications for canonical quantum gravity.  The
reinterpretation of a subset of GR solutions as solutions
of a best-matching theory in CS+V may have more interest in the
long term than the fully scale-invariant conformal gravity. This
is because the latter is much more vulnerable to experimental
disproof than GR; the CS+V theory may have a longer `shelf life'.  
Further theories which share some features with GR and conformal gravity 
are also presented, including the asymptotically flat counterparts 
of conformal gravity and the CS+V theory.

There are two important differences between the manner in which
conformal covariance is achieved in conformal gravity and the two
best known earlier attempts to create conformally covariant
theories: Weyl's 1917 theory \cite{Weyltheory} discussed in PD
and Dirac's simplified modification of it \cite{Dirac73}. First,
both of these earlier theories are spacetime theories, and their
conformal covariance leaves four-dimensional general covariance
intact. In conformal gravity and York's representation of GR, this
is not so. Second, the conformal covariance is achieved in the
theories of Weyl and Dirac through a compensating field that is
conformally transformed with the gravitational field. In Weyl's
theory, the compensating field is a 4-vector field that Weyl
identified as the electromagnetic field until Einstein \cite{Einstein1} pointed out
the difficulty discussed in PD Sec. 1. Weyl later reinterpreted the
idea of a compensating field in his effective creation of gauge
theory \cite{WeylGauge}, but he never salvaged his original
theory. In Dirac's simplification, the compensating field is the
additional scalar field in Brans--Dicke theory \cite{BD}. This
possibility has been exploited more recently in theories
with a dilatonic field \cite{Wetterich}. In
contrast, conformal gravity has no physical compensating field;
the variable $\phi$ (\ref{CoCo}) is a purely auxiliary gauge
variable used to implement conformal best matching. This is therefore 
a more radical approach, in which full scale and conformal invariance of
the gravitational field by itself is achieved.\fn{As we shall show,
one of the features of conformal gravity that ensures this is the fact
that the Lagrangian depends on \textit{ratios} of the components of the
3-metric. In checking the literature on Weyl's theory, we came across
Einstein's 1921 paper \cite{Einstein2}. In it, he follows Weyl in employing
only ratios of the 4-metric components, but drops the idea of a compensating
field. His aim is therefore similar to ours -- the gravitational field
should be made scale invariant by itself. Einstein's proposal was very tentative,
and it seems to us that it must lead to a theory with no nontrivial solutions.
We are not aware that he or anyone else attempted to take the idea further.}

We should also mention that the Lagrangian of conformal gravity
possesses \textit{two} homogeneity properties. It is homogeneous
of degree one in the velocities and of degree zero in the auxiliary
variable $\phi$. Just as scaling fixes
the potential in PD to be homogeneous of degree -2, it plays
an important role in fixing the form of conformal gravity,
especially when we include a cosmological constant (Sec. 5) and
classical bosonic matter (in Sec. 7 via general theorems proved in
Sec. 6). It turns out, as in RWR, that a universal light cone,
electromagnetism and Yang--Mills theory are enforced.  In Sec. 8,
we conclude that conformal gravity should be in agreement with the
standard tests of GR. Strong differences will emerge in cosmology
and the quantum theory; these will be less pronounced for the CS+V
theory.

\section{Summary of York's Work}

It is well known that the Hamiltonian and momentum constraints
(\ref{ham}) and  (\ref{mom}) are, respectively, the 00 and $0i$
members of the Einstein field equations (Efe's) and are
initial-value constraints. The question arises of how one can
obtain a 3-metric and associated canonical momenta that satisfy
the constraints.  A first method, which corresponds to the logic
of the best-matching procedure and uses Lagrangian variables, was
proposed by BSW in the same paper \cite{BSW} in which they
presented the BSW action (\ref{BSWac}). Freely specify $g_{ij}$,
$\dot{g}_{ij}$ and try to solve for $\xi ^{i}$. If this can be
done, the problem is solved, since the $00$ Efe in this approach
is not an equation but a square-root identity. Trying to solve the
$0i$ Efe's for $\xi^{i}$ is the thin-sandwich conjecture
\cite{TSL}. Although progress has been made, a regular method to
solve this has not been found, and counterexamples exist.

A second method is York's.  He found solutions to the constraints
by working with the Hamiltonian variables $g_{ij}$, $p^{ij}$.
Instead of using the definition of the canonical momenta \be
p^{ij} = \frac{\pa L }{\pa ({\pa {g_{ij}/\pa\lambda})}} =
\frac{\sqrt{g}}{2N}g^{ai}g^{bj}\frac{\textrm{d} g_{ab}}{\textrm{d}\lambda} \ee to
express the constraints explicitly in terms of Lagrangian
variables and the unknown $\xi^{i}$, he simply tried to find
functions $g_{ij}$ and $p^{ij}$ that satisfy (\ref{ham}) and
(\ref{mom}). In this approach, (\ref{ham}) is a proper equation
and not an identity. Indeed, what happens in York's method is that
one starts by specifying arbitrarily a pair of $3\times 3$ symmetric tensors
$\tilde{g}_{ij}$ and $\tilde{p}^{ij}$. Then there exists a regular
method to construct from $\tilde{p}^{ij}$ a $p^{ij}$ that
satisfies the momentum constraint (\ref{mom}) with respect to
$\tilde{g}_{ij}$. Once this has been achieved, one performs a
conformal transformation of $\tilde{g}_{ij}$ into $g_{ij}$ in such
a way that the new 3-metric satisfies (\ref{ham}). The end result
is that the transformed pair $g_{ij},p^{ij}$ satisfies both
constraints. The details are as follows.

In GR, an embedded hypersurface is \it maximal \normalfont if the trace \be
p^{ij}g_{ij}\equiv p = 0 \label{traceless}\ee everywhere on it. Under these
circumstances \cite{York72}, the momentum constraint (\ref{mom})
is invariant under the conformal transformation \be g_{ij}
\longrightarrow \tilde{g}_{ij} \equiv \phi^4g_{ij} , \mbox{ }
p^{ij} \longrightarrow \tilde{p}^{ij} \equiv \phi^{-4}p^{ij}, \ee
whilst the Hamiltonian constraint (\ref{ham}) becomes the
Lichnerowicz equation \cite{Lich} \be 8\nabla^2\phi + M\phi^{-7} -
R\phi = 0 , \mbox{ } gM = p_{ij}p^{ij} \geq 0.
\label{Lich} \ee The fourth power of the positive function $\phi$
(the conformal factor) is chosen to simplify the appearance of
(\ref{Lich}). York's method is to solve (\ref{mom}) in a
conformally-invariant way and then to take (\ref{Lich}) to
determine $\phi$ for $\tilde{g}_{ij}$ given.

An embedded hypersurface has \it constant mean curvature
\normalfont (CMC)\fn{The term CMC in GR refers to a constant trace
$K$ of the extrinsic curvature $K_{ab}$, which is related to the
canonical momentum by $K^{ab} = \frac{1}{\sqrt{g}}\left(p^{ab}
-\frac{p}{2}g^{ab}\right)$ and is a measure of how much the
hypersurface is bent within spacetime; we will not use $K_{ab}$ in
this paper since it is a spacetime concept.} if \be \tau =
\frac{2p}{3\sqrt{g}} = \mbox{spatial constant}. \label{Yorktime}
\ee York was able to generalize his method from maximal to CMC
hypersurfaces \cite{York72, York73}.
This generalization works because ($\ref{Yorktime}$) and
(\ref{mom}) imply that $\nabla_b(p^{ab} - \frac{p}{3}g^{ab}) = 0$
and so $\sigma^{ab} \equiv p^{ab} - \frac{p}{3}g^{ab}$ is TT
(transverse traceless), a property that York shows is invariant
under conformal transformations. Then we can find
conformally-related variables $\tilde{g}_{ij}$,
$\tilde{\sigma}^{ij}$ which obey \be
\frac{1}{\sqrt{\tilde{g}}}\left(\tilde{p}_{ab}\tilde{p}^{ab} -
\frac{1}{2}\tilde{p}^2\right) - \sqrt{\tilde{g}}\tilde{R} = 0, \ee and
the required scale factor solves the  extended Lichnerowicz equation \cite{York73} \be
8\nabla^2\phi + M\phi^{-7} -R\phi - \frac{3}{8}\tau^2\phi^5 = 0 \mbox{  }
, \mbox{  } \mbox{  } gM = \sigma_{ab}\sigma^{ab} \geq 0
\label{LichBis} 
\ee 
It is important in deriving this equation that $\tau$ and so $p/\sqrt{g}$ 
transforms as a conformal scalar: $\tilde{\tau} = \tau$.
If $\tilde{g}_{ij}$ is specified arbitrarily
and $p^{ij}$ satisfies (\ref{mom}), York was able to show that the
evolution of the conformal geometry and the scale factor $\phi$ is
uniquely determined.

Much of the strength of York's technique arises from the fact
that given a 3-metric $\tilde{g}_{ij}$ and an arbitrary
$\tilde{p}^{ij}$ there exists a well-proven method to determine
the part of $\tilde{p}^{ij}$ that is TT with respect to
$\tilde{g}_{ij}$. This solution to the first half of the problem
is not lost by the subsequent conformal transformation of
$\tilde{g}_{ij}$ determined by the solution of a Lichnerowicz
equation, which is remarkably well-behaved.      
York's method is furthermore robust to the inclusion of source
fields \cite{Niall}.  We think it significant that the only known
robust method of solving the initial-value constraints of
generally covariant GR makes no use of spacetime techniques but
rather employs 3-space conformal structures.

In GR, one regards (\ref{traceless}) and (\ref{Yorktime}) as the
maximal and CMC \it gauge conditions \normalfont respectively. The CMC
slicing yields a foliation that is extremely convenient in the
case of globally hyperbolic spacetimes with compact 3-space. The
foliation is unique \cite{unique, MT}, and the value of $\tau$
increases monotonically, either from $-\infty$ to $\infty$ in the
case of a Big-Bang to Big-Crunch cosmological solution or from
$-\infty$ to zero in the case of eternally expanding universes. In
the first case, the volume of the universe increases monotonically
from zero to a maximum expansion, at which the maximal 
condition (\ref{traceless}) is satisfied, after which it decreases
monotonically to zero. In spatially compact solutions of GR, the
total spatial volume cannot remain constant except momentarily at
maximum expansion, when $\tau \propto p =0$.  Thus, in
GR the volume is a dynamical variable.  Other authors have noticed
that these properties of $\tau$ allow its interpretation as a
notion of time, the extrinsic \it York time \normalfont
\cite{Yorktime}.

It is important to distinguish between a single initial use of the
CMC slicing condition in order to find consistent initial data and
subsequent use of the slicing when the obtained initial data are
propagated forward. This is by no means obligatory. The Efe's are
such that once consistent initial data have been found they can be
propagated with freely specified lapse and shift. This is
precisely the content of four-dimensional general covariance. If
one should wish to maintain the CMC slicing condition during the
evolution, it is necessary to choose the lapse $N$ in such a way
that it satisfies the CMC slicing equation
\be
2\left(\frac{N}{g}p_{ij}p^{ij} -\nabla^2N \right) - \frac{Np^2}{2g} = C = 
\frac{\pa}{\pa\lambda}\left(\frac{p}{\sqrt{g}}\right),
\label{LapseFixing}
\ee
where $C$ is a spatial constant.  In this paper, 
we refer to such equations as \it lapse-fixing equations\normalfont, 
because, since we do not presuppose GR, we do not always work in a 
context where the notion of slicing makes sense.  
Being homogeneous, (\ref{LapseFixing}) does
not fix $N$ uniquely but only up to global reparametrization \be
N\longrightarrow f(\lambda)N, \label{Repara}\ee where $f(\lambda)$
is an arbitrary monotonic function of $\lambda$.  Similarly, to
maintain the maximal gauge condition (\ref{traceless}), which can
only be done in the spatially non-compact case, $N$ must satisfy

\noindent the maximal lapse-fixing equation
\be
\frac{N}{g} p^{ab}p_{ab} - \nabla^2N = 0.
\label{MSE}
\ee

Now, as we show in Sec. 3, best matching w.r.t the conformal
transformations (\ref{ConfTrans}) enforces the maximal 
condition $p = 0$. We also show there that this means that in conformal gravity
the total spatial volume $V$ of the universe remains constant.
Moreover, its solutions will strongly resemble solutions of GR in
the CMC foliation at maximum expansion. The parallel with GR is
actually even closer than this. As was emphasized in PD, best
matching requires the variation w.r.t the auxiliary best-matching
variable to be performed subject to free end points. As a result,
it leads to two conditions, not one. In the case of conformal
gravity, the conformal best matching w.r.t $\phi$ leads to the
constraint $p = 0$ and to a further condition that ensures
propagation of $p=0$. The latter is identical to the CMC slicing 
equation in GR except for the modification needed to
maintain $p = 0$ rather than the marginally weaker
(\ref{Yorktime}).

In Sec. 4, we consider whether GR should be regarded as a
four-dimensionally generally covariant spacetime theory or as a
3-space theory that determines unique dynamical curves in CS+V.
Here, we merely remark that the CMC condition has hitherto been regarded
as a gauge fixing without fundamental physical significance. We
believe that our derivation in Sec. 3 of very close parallels of
both CMC conditions by conformal best matching casts new light on
this issue and supports York's contention that the two conformal
shape degrees of freedom in the 3-metric really are, together with
the volume $V$, the true dynamical degrees of freedom in spatially
compact GR.

We have already noted that there are then interpretational
differences. First, not all GR spacetimes are CMC sliceable, nor is
a CMC slicing necessarily extendible to cover the maximal
analytic extension of a spacetime \cite{SYkin}. 
Second, effects regarded as gauge artifacts in GR, such as the 
`collapse of the lapse' $N \longrightarrow 0$ in gravitational collapse 
\cite{SYkin, MT} will
be physical effects in theories in which CMC slicing has
its origin in a fundamental principle.

\section{Lagrangian Formulation of Conformal Gravity}

The basic idea of best-matching actions has been fully explained
in RWR and PD, so here we merely need to present its
application to conformal transformations. We work in the large
space Riem, which contains redundancy. To each
3-metric $g_{ij}$ on a compact 3-manifold we can apply
diffeomorphisms and conformal transformations. They generate the
infinite-dimensional orbit of $g_{ij}$, which has four dimensions
per point of the 3-manifold, three corresponding to the
diffeomorphisms and one to the conformal transformations. In the
introduction, we have already explained how the conformal
constraint takes over the role of the Hamiltonian constraint in GR
in restricting by one the degrees of freedom. The number of
physical degrees of freedom will therefore be two, as it is in GR.
Let us now see how this works out in practice.

We are looking for a consistent action along the lines of the BSW
local square-root action $S_{\mbox{\scriptsize BSW\normalsize}} =
\int \textrm{d}\lambda\int \textrm{d}^3x
\sqrt{g}\sqrt{R}\sqrt{T}$, which is additionally to be conformally
best-matched.

If we introduce the $\lambda$-dependent conformal transformation
\be
    g_{ab} \longrightarrow \omega^4 g_{ab},
\label{ConTransBis}
\ee
the velocity becomes
\be
    \frac{\textrm{d}(\omega^4g_{ab})}{\textrm{d}\lambda} \equiv 
\omega^4\left(\frac{\textrm{d}
    g_{ab}}{\textrm{d}\lambda} +
    \frac{4}{\omega}g_{ab}\left(\frac{\pa\omega}{\pa\lambda} -
    \xi^c\pa_c\omega\right)\right) \equiv
    \omega^4\left(\frac{\textrm{d}g_{ab}}{\textrm{d}\lambda} +
    \frac{4}{\omega}g_{ab}\frac{\textrm{d}\omega}{\textrm{d}\lambda}\right).
\label{Babel}
\ee

By analogy with the particle model (Sec. PD.4), we now introduce
the corrected coordinates \fn{The introduction of $\phi$ doubles
the conformal redundancy. The 3-diffeomorphism redundancy has
already been doubled by $\xi^i$. The action (\ref{SCGR}) of
conformal gravity will be defined below on $\mbox{Riem} \times
\Xi \times P$, where $\xi^i \in \Xi$ and the conformal factor
$\phi \in P$, the space of positive functions. The original
fourfold redundancy per space point of Riem is
thereby doubled.} 
\be
    \bar g_{ij}=\phi^4g_{ij}.
\label{CoCoBis}
\ee
They are trivially invariant under the pair of
compensating transformations \be
    g_{ab} \longrightarrow
    \omega^4 g_{ab},\hspace{.5cm}\phi\longrightarrow
    {\phi\over\omega},
\label{ConBanal}
\ee
which is the conformal generalization of the
banal transformation (Eq. PD.23).

Under such a transformation, the kinetic scalar
\be
    T_{\mbox{\scriptsize W\normalsize}}^{\mbox{\scriptsize C\normalsize}} =
    \phi^{-8}G^{abcd}_{\mbox{\scriptsize W\normalsize}}\frac{\textrm{d}(\phi^4g_{ab})}{\textrm{d}
    \lambda}\frac{\textrm{d}(\phi^4g_{cd})}{\textrm{d}\lambda},\label{zig}
\ee deduced from the form of (\ref{Babel}), is
invariant.\fn{Because we are using a hybrid formalism --
correcting both coordinates and velocities under conformal
transformations but only the velocities under the diffeomorphisms
-- we shall not explicitly employ barred variables in the way they
are in PD. (The reason why only the velocities need a
diffeomorphism correction is that the tensor calculus yields
the convenient scalar densities $\sqrt g$ and $\sqrt g\sqrt R$,
which are functions of the $g_{ij}$ (and their spatial derviatives)
alone. This matches the implementation of (Eq. PD.39) by
translationally invariant potentials. There are no simple
analogous invariants under (\ref{ConTransBis}), as we shall see
immediately.) Note also that the rule of
transformation of the diffeomorphism auxiliary variable under
(\ref{ConBanal}) leaves $\xi^{i}$ unchanged and changes
$\xi_{i}$.} Note that we begin with an arbitrary $W$ in the
inverse supermetric $G_{\mbox{\scriptsize W\normalsize}}^{abcd} = \sqrt{g}(g^{ac}g^{bd} -
Wg^{ab}g^{cd})$. However, whereas $W = 1$ was found in RWR
\cite{BOF} to be crucial for the consistency of GR, we will see
that this apparent freedom does not play any role in conformal
gravity.

In contrast to (\ref{zig}) and the situation in relativity
\cite{BOF}, each term in the potential part of a generalized BSW
action changes under (\ref{ConTransBis}):
\be
    \sqrt{g} \longrightarrow
    \omega^6\sqrt{g}, \mbox{ } \mbox{ } R \longrightarrow
    \omega^{-4}\left(R -
    \frac{8\nabla^2\omega}{\omega}\right),
\label{PotTransLaw} \ee with the consequence that the action of
conformal gravity must, as a `conformalization' of the BSW action,
depend not only on the velocity
$\dot{\phi}$ of the auxiliary variable $\phi$ but also on $\phi$
itself. This is what we found for the gauge variable $a$ of the
dilatation-invariant particle action (Eq. PD.24).

On the basis of (\ref{PotTransLaw}), the action for pure
(matter-free) conformal gravity is \be
    S_{\mbox{\scriptsize \normalsize}} = \int \textrm{d}\lambda    \int
    \textrm{d}^3x\left( \left(\frac{\sqrt{g}\phi^6}{V}\right)
    \sqrt{\left(\frac{\vc}{\phi^4}\right)\left( R -
    \frac{8\nabla^2\phi}{\phi}\right)} \sqrt{T^{\mbox{\scriptsize
    C\normalsize}}_{\mbox{\scriptsize W\normalsize}}
    }\right)  = \int \textrm{d}\lambda \frac { \int 
\textrm{d}^3x\sqrt{g}\phi^4 \sqrt{ R -
    \frac{8\nabla^2\phi}{\phi} } \sqrt{T^{\mbox{\scriptsize
    C\normalsize}}_{\mbox{\scriptsize W\normalsize}}  }    } {\vc },
\label{SCGR} \ee where $V$ is the `conformalized' volume \be
    V = \int \textrm{d}^3x \sqrt{g}\phi^6.\label{VolumeDef}
\ee

We first explain why the volume $V$ appears in (\ref{SCGR}).\fn{A consequence of the appearance of the 
volume in the action is that whereas the GR BSW action has intrinsic physical dimensions of length squared, 
the conformal gravity action is dimensionless (see also Sec. PD.6). The significance of this for the emergence of the equivalent notion to Newton's gravitational constant will be considered in subsequent work on the weak field limit of conformal gravity.}
Let
the Lagrangian density in (\ref{SCGR}) be $${\cal L}={\cal
L}\left(g_{ij}, {\textrm{d}g_{ij}\over\textrm{d}\lambda},
\phi,{\textrm{d}\phi\over\textrm{d}\lambda}\right),\hspace{.5cm}
S=\int\textrm{d}\lambda\int\textrm{d}^{3}x{\cal L}.$$ Provided we
ensure that $\cal L$ is a functional of the corrected coordinates
and velocities, it is bound 

\noindent to be invariant under the conformal
transformations (\ref{ConBanal}). This would be the case if we
omitted the volume $V$ in (\ref{SCGR}). But by the rules of best
matching formulated in Sec. PD.4 the Lagrangian must also be
invariant separately under all possible variations of the
auxiliary variable. Now in this case the auxiliary $\phi$ is a
function of position, so one condition that $\cal L$ must satisfy
is \be {\partial{\cal L}\over\partial\phi}=0\hspace{.2cm}
\textrm{if}\hspace{.2cm} \phi=\textrm{spatial
constant}.\label{HomCond}\ee A glance at (\ref{SCGR}) shows that
if $V$ were removed, (\ref{HomCond}) could not be satisfied. The
action must be homogeneous of degree zero in $\phi$.\fn{This is the single symmetry property that distinguishes conformal gravity that distinguishes conformal gravity from GR, which for this reason \it just \normalfont fails to be fully conformally invariant.} We exhibit
this in the first expression for $\cal L$ in (\ref{SCGR}), in
which the two expressions (\ref{PotTransLaw}) that derive from
$\sqrt g$ and $\sqrt R$ in the BSW action have been multiplied by
appropriate powers of $V$. (The second expression is more
convenient for calculations.)

A separate issue is the means of achieving homogeneity.  We
originally attempted to achieve homogeneity by using not $R^{1/2}$
but $R^{3/2}$, since $\sqrt{g}R^{3/2}$ will satisfy
(\ref{HomCond}). However, this already leads to an inconsistent
theory even before one attempts `conformalization' \cite{BOF}.
Since an ultralocal kinetic term $T$ has no conformal weight,
another possibility would be to construct a conformally-invariant
action by multiplying such a $T$ by a conformally-invariant
3-dimensional scalar density of weight 1. Whereas the Bach tensor
and the Cotton--York tensor are available conformally-invariant
objects \cite{York71}, the suitable combinations they give rise to
are cumbersome, and much hard work would be required to
investigate whether any such possibilities yield consistent
theories. Even if they do, they will certainly be far more
complicated than conformal gravity.

We have therefore simply used powers of the volume $V$, which has
precedent in the variational problems associated with the
Yamabe conjecture \cite{Yamabe}. We believe that the resulting
theory, conformal gravity, is much simpler than any other
potentially viable theory of scale-invariant gravity. The use of
the volume has the added advantage that $V$ is conserved. This
ensures that conformal gravity shares with the particle model the
attractive properties it acquires from conservation of the moment
of inertia $I$.  We extend this method (of using powers of $V$ to
achieve homogeneity) in Secs. 6 and 7 to include matter coupled to
conformal gravity. The consequences of this are discussed in Sec.
8. It is the use of $V$ that necessitates our assumption of a
spatially compact manifold $\cal M$. It is a physical assumption,
not a mere mathematical convenience. It would not be necessary in the
case of theories of the type considered in the previous paragraph.

We must now find and check the consistency of the equations of
conformal gravity. The treatment of best matching in the particle
model in Sec. PD.4 tells us that we must calculate the canonical
momenta of $g_{ij}$ and $\phi$, find the conditions that ensure
vanishing of the variation of the Lagrangian separately w.r.t to
possible independent variations of the auxiliary $\phi$ and its
velocity $v_{\phi}=\textrm{d}\phi/\textrm{d}\lambda$, and then
show that these conditions, which involve the canonical momenta,
together with the Euler--Lagrange equations for $g_{ij}$ form a
consistent set. This implements best matching by the
free-end-point method (Sec. PD.4).

The canonical momentum $p_{\phi}$ of $\phi$ is 
\be
p_{\phi}\equiv{\partial{\cal L}\over\partial v_{\phi}} = 
\frac{\sqrt{g}}{2N\vc}G_{\mbox{\scriptsize W\normalsize}}^{abcd}\frac{\textrm{d}(\phi^4g_{ab})}
{\textrm{d}\lambda}\frac{4}{\phi}g_{cd},
\label{confcanmom} 
\ee 
where 
$2N = \sqrt{\frac{T^{\mbox{\scriptsize C\normalsize}}_
{\mbox{\scriptsize W\normalsize }}}{R - \frac{8\nabla^2\phi}{\phi}}}$ 
is (twice) the `lapse' of matter-free conformal gravity. The gravitational
canonical momenta are 
\be 
p^{ab} =
\frac{\sqrt{g}}{2N\vc}G_{\mbox{\scriptsize W\normalsize}}^{abcd}\frac{\textrm{d}(\phi^4g_{cd})}{\textrm{d}\lambda},
\label{gravcanmom} \ee 
and we see that the canonical momenta
satisfy the primary constraint 

\noindent
\be
p = \frac{\phi}{4}p_{\phi},
\label{primconstr} \ee where \be p=p^{ab}g_{ab}\label{Trace}\ee is
the trace of $p^{ab}$ and is the `localized' analogue of the
dilatational momentum (Eq. PD.3). The primary constraint (\ref{primconstr}) is a
direct consequence of the invariance of the action (\ref{SCGR})
under the banal transformations (\ref{ConBanal}).

Independent variations $\delta\phi$ and $\delta v_{\phi}$ of
$\phi$ and its velocity in the instantaneous Lagrangian that can
be considered are: 1) $\delta\phi$ is a spatial constant and
$\delta v_{\phi}\equiv 0$; 2) $\delta\phi$ is a general function
of position and $\delta v_{\phi}\equiv 0$; 3) $\delta\phi\equiv 0$
and $\delta v_{\phi}\neq 0$ in an infinitesimal spatial region.
The possibility 1) has already been used to fix the homogeneity of
$\cal L$. Let us next consider 3). This tells us that
$\partial{\cal L}/\partial v_{\phi}=0$. But $\partial{\cal
L}/\partial v_{\phi}\equiv p_{\phi}$, so we see that the canonical
momentum of $\phi$ must vanish. Then the primary constraint
(\ref{primconstr}) shows that \be p=0.\label{Pisa}\ee

As a result of this, without loss of generality, we can set $W$, the 
coefficient of the trace, equal to zero, $W = 0$, 
in the generalization of the DeWitt supermetric used in 
(\ref{SCGR}). If conformal gravity proves to be a viable theory,
this result could be significant, especially for the quantization
programme (see Sec. 8), since it ensures that conformal gravity,
in contrast to GR, has a positive-definite kinetic energy. Indeed, in GR,
since $$ \frac{\pa g_{ab}}{\pa\lambda} =
\frac{2N}{\sqrt{g}}\left(p_{ab} - \frac{p}{2}g_{ab}\right) + 2
\nabla_{(a}\xi_{b)} $$
the rate of change of $\sqrt g$, which defines the volume element
$\sqrt g\textrm{d}^3x$, is measured by the trace $p$,
$$\frac{\pa\sqrt{g}}{\pa\lambda} = - \frac{Np}{2} + \sqrt{g}\nabla_a\xi^a.$$
Therefore, in conformal gravity the volume element -- and with it the
volume of 3-space -- does not change and cannot make a contribution to the 
kinetic
energy of the opposite sign to the contribution of the shape degrees
of freedom.  Finally, we consider 2) and (including the use of $p_{\phi} = 0$) we obtain 
\be
    \phi^3N\left(R - \frac{7\nabla^2\phi}{\phi}\right) -
    \nabla^2(\phi^3N) = \phi^5\left<\phi^4N \left( R -
    \frac{8\nabla^2\phi}{\phi} \right) \right>,
\label{fullslicing} 
\ee
where we use the usual notion of global average: 
\be
    <A> = \frac{\int
    \textrm{d}^3x \sqrt{g} A} {\int \textrm{d}^3x\sqrt{g} }.
\ee
This lapse-fixing equation holds at all times.  Thus it has a status different from the constraints; it has no analogue
in the particle model in PD or any other gauge theory of which
we are aware.   It is, however, a direct consequence of conformal
best matching, and as explained below, it plays an important role in the mathematical structure of conformal gravity.  

Besides the trace constraint (\ref{Pisa}), $p^{ab}$ must satisfy
the primary quadratic constraint \be
    -{\cal H}^{\mbox{\scriptsize C\normalsize}} \equiv
    \frac{\sqrt{g}\phi^4}{V^{\frac{2}{3}}}\left( R -
    \frac{8\nabla^2\phi}{\phi} \right)
    - \frac{\vc}{\sqrt{g}\phi^4}p^{ab}p_{ab} = 0
\label{fullham} \ee due to the local square-root form of the
Lagrangian, and the secondary linear constraint \be
    \nabla_bp^{ab} = 0
\label{cmom} \ee from variation with respect to $\xi_{i}$. Of
course, (\ref{cmom}) is identical to the GR momentum constraint
(\ref{mom}), while (\ref{fullham}) is very similar to the
Hamiltonian constraint (\ref{ham}).

The Euler--Lagrange equations for $g_{ij}$ are
\be
\begin{array}{ll}
\frac{\mbox{\scriptsize d\normalsize}p^{ab}}{\mbox{\scriptsize d\normalsize}\lambda} = & 
\frac{\phi^4\sqrt{g}N}{\vc}\left(Rg^{ab} - \frac{4\nabla^2\phi}{\phi} - R^{ab}\right) 
- \frac{\sqrt{g}}{\vc}(  g^{ab}\nabla^2(\phi^4N) - \nabla^a\nabla^b(\phi^4N)  ) 
- \frac{2N\vc}{\sqrt{g}\phi^4}p^{ac}{p^b}_c 
\\ &
+ \frac{8\sqrt{g}}{\vc}
\left(
\frac{1}{2}g^{ab}g^{cd} - g^{ac}g^{bd}
\right)
(\phi^3N)_{,c}\phi_{,d}
- \frac{2\sqrt{g}}{3\vc}\phi^6g^{ab}
\left<
\phi^4N
\left( 
R - \frac{8\nabla^2\phi}{\phi}
\right)
\right> 
+ \frac{4}{\phi}p^{ab}\frac{\mbox{\scriptsize d\normalsize}\phi}{\mbox{\scriptsize d\normalsize}\lambda} .
\end{array}\label{GravEL}
\ee 
They can be used to check the consistency of the full set of
equations, constraints and lapse-fixing equation of conformal gravity. To
simplify these calculations and simultaneously establish the
connection with GR, we go over to the distinguished representation
(Sec. PD.4), in which $\phi = 1$ and $\xi^i = 0$. The three
constraints that must be satisfied by the gravitational canonical
momenta are
\be
    -{\cal H}^{\mbox{\scriptsize C\normalsize}} \equiv
    \frac{\sqrt{g}}{V^{\frac{2}{3}}}R - \frac{\vc}{\sqrt{g}}p^{ab}p_{ab} =
    0,\hspace{.5cm}\nabla_bp^{ab} = 0,\hspace{.5cm}p=0.
\label{cham}
\ee
The lapse-fixing equation (\ref{fullslicing}) becomes\fn{We can replace $R$
by $\frac{V^{\frac{4}{3}}}{g}p^{ab}p_{ab}$ in this expression
by use of the Hamiltonian constraint, thus bringing the appearance
of (\ref{BMFLFC}) in accord with (\ref{LapseFixing}) and (\ref{MSE}). The 
form
(\ref{BMFLFC}) is more convenient for later use.}
\be
\nabla^2N - NR = - <NR>, \label{BMFLFC}
\ee
and the Euler--Lagrange
equations are
\be
\begin{array}{ll}
    \frac{\textrm{\scriptsize{d}}p^{ab}}{\textrm{\scriptsize{d}}\lambda} = &
    \frac{\sqrt{g}N}{\vc}(Rg^{ab} - R^{ab}) -
    \frac{\sqrt{g}}{\vc}(  g^{ab}\nabla^2N -
    \nabla^a\nabla^bN  ) - \frac{2N\vc}{\sqrt{g}}p^{ac}{p^b}_c -
    \frac{2\sqrt{g}}{3\vc}g^{ab}<NR>.
\end{array}\label{BMFGravEL}
\ee

Since the volume $V$ is conserved, and its numerical value depends
on a nominal length scale, for the purpose of comparison with the
equations of GR we can set $V=1$. (This cannot, of course, be done
before the variation that leads to the above equations, since the
variation of $V$ generates forces. It is important not to confuse
quantities on-shell and off-shell.) We see, setting $V=1$, that
the similarity with GR in York's CMC slicing is strong.  In fact, 
the constraints (\ref{cham}) are identical to the GR constraints and 
York's slicing condition at maximal expansion, and
the Euler--Lagrange equations differ only by the absence of the GR
term proportional to $p$ and by the presence of the
final force term. This has the same form as the force due to
Einstein's cosmological constant, but its strength is fixed by the
theory, just as happens for the strength (Eq. PD.39) of the induced
cosmological force in the particle model. Finally, there is the
lapse-fixing equation (\ref{BMFLFC}), which is an eigenvalue equation of
essentially the same kind as the lapse-fixing equation
(\ref{LapseFixing}) required to maintain York's CMC slicing
condition (\ref{Yorktime}). In accordance with the last footnote,
the left-hand sides are identical, and
in both cases the spatial constant on the right-hand side is
determined by the functions of position on the left-hand side. 

In our view, one of the most interesting results of
this work is the derivation of such lapse-fixing equations directly from a
fundamental symmetry requirement rather than as an equation which could 
be interpreted as maintaining a gauge fixing.  We now develop this point.  

This is part of the confirmation that we do have a consistent set
of equations, constraints and lapse-fixing equation.  We show this in Sec. 4,
which amounts to demonstrating that if the constraints (\ref{cham}) hold 
initially, then they will propagate due to the Euler--Lagrange equations (\ref{BMFGravEL})
\textit{and} the lapse-fixing equation (\ref{fullslicing}).  
The propagation of the vector momentum constraint is always
unproblematic, being guaranteed by the 3-diffeomorphism invariance
of the theory. The two constraints that could give difficulty are
the quadratic and linear scalar constraints. In RWR \cite{BOF},
the propagation of the quadratic constraint proved to be a
delicate matter and generated the new results of that paper.
However, in this paper we are merely `conformalizing' the results
of \cite{BOF}, and we shall see that the consistency
achieved for the quadratic constraint in \cite{BOF} carries
forward to conformal gravity. The only issue is therefore whether
the new constraint $p=0$ is propagated.  Now, we find that the form of 
$\dot{p}$ evaluated from the Euler--Lagrange equations \it is automatically guaranteed to be zero \normalfont 
by virtue of the lapse-fixing equation.  Thus the propagation of $p = 0$ is guaranteed rather than separately imposed.  
It is in this sense that conformal gravity is not maintaining a gauge fixing.  

We conclude this section with a brief discussion of the
thin-sandwich problem (TSP) for conformal gravity. As we said at
the start of Sec. 2, the TSP is a serious issue for GR in the
Lagrangian formalism. We have a different, not necessarily easier
TSP to investigate in conformal gravity.  In both cases, we can
bypass the TSP by working instead in the Hamiltonian formalism
(Sec. 4). It is however worth stating the problem, counting the
freedoms and the conditions that should fix them, and making some
comments.

In the \textit{conformal thin-sandwich problem}, one specifies
$g_{ij}$ and $\textrm{d}{g}_{ij}/\textrm{d}\lambda$ (12 numbers
per space point) and uses the momentum constraint, the trace
constraint and the lapse-fixing equation (all expressed in
Lagrangian variables) to find $\phi$ and $\xi^i$ that make
$p_{ij}$ TT w.r.t the best-matched corrected coordinate $\phi^4
g_{ij}$, where $\phi$ is part of the solution of the TSP. To solve
this problem we have five equations. This is the correct number
that we need; four to impose a TT condition just as in York's
Hamiltonian method and the fifth to move $g_{ij}$ along its orbit
to the best-matched position, which reflects the action of the
Lichnerowicz equation (\ref{Lich}).\fn{The best matching for
conformal gravity thus differs characteristically from the
3-diffeomorphism best matching, which corrects the `direction of
the velocity' in the same way at any point on the orbit in Riem.
In the conformal problem, both the direction of the velocity and
the position on the orbit are corrected. Many relativists find the
need for and effect of the Lichnerowicz transformation puzzling. We believe
this and the next section provide a transparent geometrical
(best-matching) explanation of the transformation.} The count of freedoms 
works
out too. Among the 12 numbers per space point initially specified,
two determine a position in CS and two a direction in CS. These
are the physical degrees of freedom, and they are unchanged by the
solution of the TSP. Three numbers must remain free, since they
correspond to the freedom to specify the coordinates on the
manifold freely. The remaining five are corrected by the five
conditions per space point.

\section{Hamiltonian Formulation and Alternative Theories}

In this section, we examine the Hamiltonian formulation.  We exercise a number of options in constructing 
theories such as GR and conformal gravity.  In particular, while we have hitherto considered only compact spaces without boundary, we now also touch on the asymptotically-flat case.  We also attempt to incorporate both maximal 
and CMC conditions.  Our methods of doing so necessitate a discussion of whether the lapse-fixing equations are fundamental or a gauge-fixing.  When we conformally-correct both $\dot{g}_{ij}$ and $g_{ij}$ (which corresponds below to the use of two auxiliary variables which we later find are related), we find that the lapse-fixing equations are 
variationally guaranteed.  This is the case in conformal gravity.  
When this occurs, the alternative gauge-fixing interpretation is not available.
With the hindsight of the last section, one must apply this full correction to obtain the most complete 
theories.  We build up the Hamiltonian structure toward this completeness.  

We begin with the case that corresponds to working on superspace.    
ADM and Dirac showed that the Hamiltonian for GR can be written as
(\ref{ADM}) -- (\ref{mom}):
\be
\mbox{\sffamily H\normalfont} =
\int\left(N\left(\frac{1}{\sqrt{g}}\left(p^{ab}p_{ab} - {p^2 \over 2}\right)
- \sqrt{g}R\right) +
\xi^i\left(-2
\nabla_j{p^j}_i\right)\right) \textrm{d}^3x.\label{Ham}
\ee
Hamilton's equations are then the
evolution equations
\be
\begin{array}{l}
\frac{\mbox{\scriptsize d\normalsize} g_{ab}}{\mbox{\scriptsize d\normalsize}\lambda} = 
\frac{2N}{\sqrt{g}}\left(p_{ab} - \frac{p}{2}g_{ab}\right),
\label{smallevol}  
\end{array}
\ee
\be
\begin{array}{l}
\frac{\mbox{\scriptsize d\normalsize} p^{ab}}{\mbox{\scriptsize d\normalsize}
\lambda} = \sqrt{g}N
\left(
\frac{R}{2}g^{ab} - R^{ab}
\right)
- {\sqrt{g}}(  g^{ab}\nabla^2N - \nabla^a\nabla^bN  ) 
+ \frac{N}{2\sqrt{g}}g^{ab}\left(p^{ij}p_{ij} - \frac{p^2}{2}\right) 
- \frac{2N}{\sqrt{g}}\left(p^{ac}{p^b}_c - \frac{p}{2}p^{ab}\right).
\label{BSWEL}
\end{array}
\ee
Here, $N$ and $\xi^i$ are arbitrary and the evolution equations propagate the 
constraints. These are equivalent to the BSW evolution equations of RWR.

Now we wish to work with CMC slices on CS+V.  In this case we find that 
there is little difference between the compact and asymptotically-flat cases.  
We can treat this in the Hamiltonian framework simply by adding 
another constraint to the Hamiltonian (\ref{Ham}), i.e by considering 
\be
\mbox{\sffamily H\normalfont}^{\eta} 
= \int \textrm{d}^3x (N{\cal H} + \xi^i{\cal H}_i + (\nabla^2\eta)p)
\label{ham'}
\ee
and treating $\eta$ as another Lagrange multiplier.  The Laplacian is introduced here to obtain the CMC condition 
as the new constraint arising from $\eta$-variation: $\nabla^2p = 0 \Rightarrow {p}/{\sqrt{g}} = \mbox{const.}$.  
In addition we get the standard GR Hamiltonian and momentum constraints from variation w.r.t the Lagrange
multipliers $N$ and $\xi^i$.  Now, we can impose $\frac{\textrm{\scriptsize{d}}}{\textrm{\scriptsize{d}}\lambda}\left(\frac{p}{\sqrt{g}}\right) = const$ 
to arrive at the CMC lapse-fixing equation.  Whereas one could reinterpret this as the study in the CMC gauge of 
the subset of (pieces of) GR solutions which are CMC foliable, we can also consider this to be a new theory with a preferred 
fundamental CMC slicing.\fn{By a preferred fundamental slicing, we mean a single stack 
of Riemannian 3-spaces.  This is not to be confused with GR, where there are an infinity of such stacks between any two given spacelike hypersurfaces, 
each of which is called a slicing or foliation of that spacetime region.}    
The latter interpretation is our first CS+V theory.  

This CS+V theory's evolution equations are (\ref{smallevol}) and (\ref{BSWEL}) but picking up the 
extra terms $g_{ab}\nabla^2\eta$ and $-p^{ab}\nabla^2\eta$ respectively.  
We have already dealt with the CMC constraint;     
it turns out that we need to set $\nabla^2\eta = 0$ to preserve the other constraints.
Therefore the CMC Hamiltonian is well-defined when one makes the gauge choices of 
$\nabla^2\eta = 0$ and that $N$ satisfies the CMC lapse-fixing equation.
     
It is more satisfactory however to introduce a second auxiliary variable to 
conformally correct the objects associated with the metric.  
We use $(1 + \nabla^2\zeta)^{\frac{1}{6}}$ in place of $\phi$ since this implements 
volume-preserving conformal transformations 
\be
\bar{V} - V = \int \textrm{d}^3x\sqrt{g}\left( (1+ \nabla^2\zeta)^{\frac{1}{6}}\right)^6 - \int \textrm{d}^3x\sqrt{g} = \int \textrm{d}^3x\sqrt{g}\nabla^2\zeta = \oint \textrm{d}S_a\nabla^a\zeta = 0 
\ee
in the case where the 3-geometry is closed without boundary.  Applying these corrections, we obtain 

\noindent the Hamiltonian 
\be
\mbox{\sffamily H\normalfont}^{\zeta\eta} 
= \int \textrm{d}^3x (N{\cal H}^{\zeta} + \xi^i{\cal H}_i + (\nabla^2\eta)p)\label{dcham'},
\ee
\be 
{\cal H}^{\zeta} \equiv  \frac{1}{\sqrt{g}}\left(\frac{\sigma_{ij}\sigma^{ij}}{(1 + \nabla^2\zeta)^{\frac{2}{3}}} - \frac{(1+ \nabla^2\zeta)^{\frac{4}{3}}p^2}{6}\right) 
 - \sqrt{g}(1 + \nabla^2\zeta)^{\frac{2}{3}}\left(R - \frac{8\nabla^2(1 + \nabla^2\zeta)^{\frac{1}{6}}}{(1 + \nabla^2\zeta)^{\frac{1}{6}}}\right). 
\ee
Note that the conformal weighting of $\sigma_{ij}\sigma^{ij}$ and $p^2$ is distinct, just as in York's method!  We require 
$p$ to transform as a conformal scalar.  Now, 
$N$, $\xi^i$, $\eta$ variation yields ${\cal H}^{\zeta} = 0$, ${\cal H}_i = 0$, $p/{\sqrt{g}} = const$. 
Then in the distinguished representation $\nabla^2\eta = \nabla^2\zeta = 0$, 
$\zeta$-variation and one use of ${\cal H}^{\zeta} = 0$ yields
\be
NR - \nabla^2N + \frac{Np^2}{4g} = const.
\ee 
Now, this is of the correct form to automatically guarantee that $\frac{\textrm{\scriptsize{d}}}{\textrm{\scriptsize{d}}\lambda}\left(\frac{p}{\sqrt{g}}\right) = const$.  
Thus the CMC lapse-fixing equation is variationally encoded in this \it full CS+V theory\normalfont.  Furthermore, ${\cal H}^{\zeta} = 0$ is the 
extended Lichnerowicz equation (\ref{LichBis}) for $\phi = 1 + \nabla^2\zeta$.  It is this equation that carries the auxiliary variables 
which encode the CMC lapse-fixing equation.  
   
We now try to extend the above workings to the case of the maximal condition.  This involves starting with a slightly different Hamiltonian,   
\be
\mbox{\sffamily H\normalfont}^{\theta} = \int \textrm{d}^3x (N{\cal H} +
\xi^i{\cal H}_i - \theta p).
\label{pooras}
\ee  
Variation w.r.t $\theta$ gives us the maximal condition $p = 0$.  
In the asymptotically-flat case, it makes sense to impose $\dot{p} = 0$ to arrive at the maximal lapse-fixing  equation
\be
\nabla^2N = RN,
\label{MSE2}
\ee
since this is an extremely well-behaved equation when taken in conjunction with the
boundary condition $N \longrightarrow 1$ at infinity.  Again, variation w.r.t the 
multipliers $N$, $\xi^i$ yields the standard GR Hamiltonian and momentum constraints 
respectively.  Now if we choose $\theta = 0$ in the evolution we
preserve the Hamiltonian and momentum constraints.   
Therefore $\mbox{\sffamily H\normalfont}^{\theta}$ can be interpreted as 
yielding standard GR with a choice of maximal gauge, or alternatively that this 
represents a new theory with a preferred fundamental maximal slicing: our first asymptotically-flat theory.

This does not work in the case of a compact manifold without boundary, since now the only solution of (\ref{MSE2}) is $N \equiv 0$ and so we get frozen dynamics.  
But if we use the volume of the universe (which amounts to moving away from GR), we are led to new theories.  
First, consider
\be
\mbox{\sffamily H\normalfont}^{\theta \mbox{\scriptsize V\normalsize}} 
= \int \textrm{d}^3x(N{\cal H}^{\mbox{\scriptsize V\normalsize}} + \xi^i{\cal H}_i - \theta p) \mbox{ }, \mbox{  }
{\cal H}^{\mbox{\scriptsize V\normalsize}} \equiv
\frac{V^{\frac{2}{3}}}{\sqrt{g}}p^{ab}p_{ab}
- \frac{\sqrt{g}}{V^{\frac{2}{3}}} R 
\label{poormanscg}
\ee
Then the $N$, $\xi^i$, $\theta$ variations yield the constraints ${\cal H}^V = 0$,  ${\cal H}_i = 0$ and $p = 0$.  
Imposing $\dot{p} = 0$ yields a maximal lapse-fixing equation 
\be
\nabla^2N = RN - <RN>.  
\label{nav}
\ee
Whereas this could be regarded as a gauge fixing, the underlying theory is no longer GR.  
The other interpretation is that of a new theory with a preferred fundamental maximal slicing.  

Second, consider the use of two auxiliary conformal variables:  

\be
\mbox{\sffamily H\normalfont}^{\phi\theta \mbox{\scriptsize V\normalsize}} = \int \textrm{d}^3x
(N{\cal H}^{\mbox{\scriptsize C\normalsize}} + \xi^i{\cal H}_i - \theta p) \mbox{ },\mbox{ } 
{\cal H}^{\mbox{\scriptsize C\normalsize}} \equiv
\frac{V^{\frac{2}{3}}}{\phi^4\sqrt{g}}p^{ab}p_{ab}
- \frac{\phi^4\sqrt{g}}{V^{\frac{2}{3}}} \left(R - \frac{8\nabla^2\phi}{\phi}\right).  
\label{cghamiltonian}
\ee
Then the $N$, $\xi^i$, $\theta$ variations yield the constraints 
${\cal H}^{\mbox{\scriptsize C\normalsize}} = 0$,  ${\cal H}_i = 0$ and $p = 0$.  
Hamilton's equations are now the evolution equations

\noindent
\be
\begin{array}{ll}   
\frac{\textrm{\scriptsize{d}} g_{ab}}{\textrm{\scriptsize{d}}\lambda} =
&\frac{2N\vc}{\sqrt{g}}p_{ab} - \theta g_{ab},\cr
\frac{d p^{ab}}{d \lambda} = &
\frac{\sqrt{g}N}{\vc}\left(\frac{R}{2}g^{ab} - R^{ab}\right)
- \frac{\sqrt{g}}{\vc}(  g^{ab}\nabla^2N - \nabla^a\nabla^bN  )
\\ &    +
\frac{N\vc}{2\sqrt{g}}g^{ab}p^{ij}p_{ij}  
    -
\frac{2N\vc}{\sqrt{g}}p^{ac}{p^b}_c
    - \frac{2}{3\vc}\left<NR\right>g^{ab} +
\theta p^{ab}
\end{array}
\label{BSWELMOD}
\ee
in the distinguished representation $\phi = 1$.  
Whilst $\phi$ variation yields (\ref{nav}), 
Hamilton's equations give 
\be
\dot{p} =  \frac{2\sqrt{g}}{\vc}(NR - <NR> - \nabla^2N)
\ee
which is thus automatically satisfied due to (\ref{nav}).  
This theory is (full) conformal gravity, as can be confirmed by 
Legendre transformation and BSW elimination \cite{BSW} 
to recover the Lagrangian of Sec. 3.  A nontrivial step in this 
procedure of justifying the Hamiltonian (\ref{cghamiltonian}) to be that of conformal gravity  
is presented 
shortly.      
Because the lapse-fixing equation that maintains the condition 
is also guaranteed, conformal gravity is definitely \it not \normalfont 
interpretable as a gauge fixing.  

Before turning to these, we mention that there is one further maximal theory which arises from considering two conformal auxiliary variables in 
the asymptotically flat case:
\be
\mbox{\sffamily H\normalfont}^{\phi\theta} = \int \textrm{d}^3x
(N{\cal H}^{\phi} + \xi^i{\cal H}_i - \theta p) \mbox{ },\mbox{ } 
{\cal H}^{\phi} \equiv
\frac{1}{\phi^4\sqrt{g}}p^{ab}p_{ab}
- {\phi^4\sqrt{g}} \left(R - \frac{8\nabla^2\phi}{\phi}\right).  
\label{ascg}
\ee 
Then $N$, $\xi^i$, $\theta$ variation yield the constraints ${\cal H}^{\phi} = 0$,  ${\cal H}_i = 0$ and $p = 0$.  
Now, $\phi$ variation gives the maximal lapse-fixing equation (\ref{MSE2}), and so automatically guarantees the propagation of 
the condition $p = 0$.  This is our \it full asymptotically flat theory\normalfont.  
Note that ${\cal H}^{\phi} = 0$ is the Lichnerowicz equation (\ref{Lich}).  Again, the Lichnerowicz equation carries the auxiliary variable 
which encode the lapse-fixing equation.  For conformal gravity itself, we see it as significant that one must modify the corresponding Lichnerowicz equation 
by the introduction of volume terms to get an analogous scheme.  

Like conformal gravity, our full asymptotically-flat theory has no role for $p$ in its dynamics, but, unlike conformal gravity, it does not possess global terms.  
We are less interested in this theory than in conformal gravity because it would not be immediately 
applicable to cosmology, on account of being asymptotically flat.  These two theories should be contrasted with our full CS+V theory, in which $p$ does play a role, 
which means that the standard GR explanation of cosmology is available.  We believe the CS+V theory merits a full treatment elsewhere as another potential rival to GR.

The theories (\ref{ham'}), (\ref{pooras}, (\ref{poormanscg}) above may be viewed as a poor man's versions of our three full theories.  
The formulation provided by their gauge interpretation is new and may be useful in numerical relativity.  
  
As in the Appendix of PD, we now justify the Hamiltonians of GR and conformal gravity (which are the main theories discussed in this paper) 
from the point of view of free-end-point variation.  Lanczos \cite{qm} treats the Hamiltonian method 
as a special case of the Lagrangian method for which the kinetic energy has the form $T = \Sigma_ip^i\dot{q}_i$.  Thus we treat the momenta $p^i$ as coordinates 
on an equal footing with the $q_i$.  Constraints are to be appended with Lagrange multipliers.  
We furthermore treat all the variables that we regard as gauge auxiliaries by the free-end-point method.  

For GR, this includes treating the shift as the velocity of some auxiliary variable $f^i$: $\xi^i = \dot{f}^i$.  
Thus
\be 
L = \int \textrm{d}^3x{\cal L} = \int \textrm{d}^3x (p^{ij}\dot{g}_{ij} + p^{\mbox{\scriptsize f\normalsize}}_i\dot{f}^i - N{\cal H} 
- S^i(p^{\mbox{\scriptsize f\normalsize}}_i - 2\nabla_j{p^j}_i)),
\ee
there is by construction no $f^i$ variation and $\dot{f}^i$ variation yields $p^{\mbox{\scriptsize f\normalsize}}_i = 0$.  
Variation w.r.t the true multipliers $N$ and $S_i$ yields respectively the Hamiltonian constraint ${\cal H} = 0$ and 
$-2\nabla_j{p^j}_i = -p^{\mbox{\scriptsize f\normalsize}}_i$ (i.e the momentum constraint ${\cal H}_i = 0$).  
Variation w.r.t $p^{\mbox{\scriptsize f\normalsize}}_i$ yields $S^i = \dot{f}^i = \xi^i$, thus recovering the habitual Lagrange  multiplier notion for the shift, 
and variation w.r.t $p^{ij}$ and $g_{ij}$ yields the usual ADM evolution equations.  
Thus we can write $L =\int \textrm{d}^3x(p^{ij}\dot{g}_{ij} - N{\cal H} - \xi^i{\cal H}_i)$, so it is consistent for 
the Hamiltonian to take its usual GR form, $\mbox{\sffamily H\normalfont} = \int \textrm{d}^3x(N{\cal H} + \xi^i{\cal H}_i)$.

For conformal gravity,
\be
L = \int \textrm{d}^3x\left(p^{ij}\dot{g}_{ij} + p^{\mbox{\scriptsize f\normalsize}}_i\dot{f}^i +p_{\phi}\dot{\phi} 
- N{\cal H}^{\mbox{\scriptsize C\normalsize}} - S^i(p^{\mbox{\scriptsize f\normalsize}}_i 
- 2\nabla_j{p^j}_i) - \theta\left(\frac{\phi}{4}p_{\phi} - p^{ij}\dot{g}_{ij}\right)\right),
\label{longlag}
\ee
Repeat all the GR steps (obtaining ${\cal H}^{\mbox{\scriptsize C\normalsize}} = 0$ and the conformal gravity evolution equations in place of their GR ADM counterparts).  In addition we have a new auxiliary variable $\phi$; 
variation w.r.t this yields the lapse-fixing equation (\ref{nav}) and $\dot{\phi}$ variation yields $p_{\phi} = 0$.  
We also have a new true multiplier $\theta$ variation w.r.t which yields $p = \frac{\phi}{4}p_{\phi} = 0$.  
The virtue of (\ref{longlag}) is that $p_{\phi}$ variation directly yields $\theta = 4\frac{\dot{\phi}}{\phi}$.
Thus we can write $L = \int \textrm{d}^3x(p^{ij}\dot{g}_{ij} - N{\cal H}^{\mbox{\scriptsize C\normalsize}} - \xi^i{\cal H}_i + \theta p)$, so it is consistent to take the conformal gravity Hamiltonian to be 
$\int \textrm{d}^3x (N{\cal H}^{\mbox{\scriptsize C\normalsize}} + \xi^i{\cal H}_i - \theta p)$.   

We now consider the preservation of the other conformal gravity constraints ${\cal H}^{\mbox{\scriptsize C\normalsize}}$ and ${\cal H}_i$.
We find that we need to set $\theta = 0$.
Therefore (by comparison with GR) the only term we need to worry about is the
$-\frac{2}{3\vc}<NR>g^{ab}$ term in the
$\dot{p}^{ab}$ equation.
Since it is of the form $Cg^{ab}$ it clearly will not
disturb the momentum constraint.  Therefore we need only worry about
conserving the Hamiltonian constraint.  This is quite
straightforward.  We first realize that
\be
\frac{\pa\sqrt{g}}{\pa\lambda} = \sqrt{g}\nabla_i\xi^i.
\label{nog}
\ee
Hence
\be
\frac{\pa V}{\pa \lambda} = 0
\label{nov}.
\ee
The only other term to worry about is
$\frac{\vc}{\sqrt{g}}p_{ab}p^{ab}$. Varying $p^{ab}$ gives
$-\frac{4}{3\sqrt{g}}<NR>p = 0$. Therefore the constraints are
preserved under evolution.  

We now show that we can just as easily treat the Hamiltonian dynamics of conformal gravity in the general
representation.  The lapse-fixing equation from $\phi$ variation is now 
\be
\nabla^2(\phi^3N) = \phi^3N\left(R - \frac{7\nabla^2\phi}{\phi}\right) - \phi^5\left<\phi^4N \left( R - \frac{8\nabla^2\phi}{\phi} \right) \right>,
\label{fullslicing2} 
\ee
whilst Hamilton's evolution equations are now 
\be
\begin{array}{ll}   
\frac{\textrm{\scriptsize{d}} g_{ab}}{\textrm{\scriptsize{d}}\lambda} 
= &\frac{2N\vc}{\phi^4\sqrt{g}}p_{ab} - \theta g_{ab},\cr
\frac{d p^{ab}}{d \lambda} = & 
\frac{\phi^4\sqrt{g}N}{\vc}\left(\frac{R}{2}g^{ab} - R^{ab}\right) 
+ \frac{N\vc}{2\phi^4\sqrt{g}}p^{cd}p_{cd}g^{ab}  
- \frac{\sqrt{g}}{\vc}(g^{ab}\nabla^2(\phi^4N) - \nabla^a\nabla^b(\phi^4N)  ) 
+ \theta p^{ab} \\&      
-\frac{2N\vc}{\sqrt{g}\phi^4}p^{ac}{p^b}_c  
+ \frac{8\sqrt{g}}{\vc}
\left(
\frac{1}{2}g^{ab}g^{cd}-g^{ac}g^{bd}
\right)
(\phi^3N)_{,c}\phi_{,d}
- \frac{2\sqrt{g}}{3\vc}\phi^6g^{ab}
\left<
\phi^4N
\left( 
R\mbox{--}\frac{8\nabla^2\phi}{\phi}
\right)
\right> 
  .\label{ham1}
\end{array}
\ee
We can compare these expressions to their Lagrangian analogues (\ref{gravcanmom}), (\ref{GravEL})
and we see they coincide if $\theta = \frac{4}{\phi}\frac{\textrm{\scriptsize{d}}\phi}{\textrm{\scriptsize{d}}\lambda}$.
This
will guarantee that the constraints are preserved by the evolution.
Alternatively, we could evolve the constraints using the
Hamiltonian evolution
equations (\ref{ham1}) and discover that the
constraints propagate if and only
if the lapse function, $N$,
satisfies (\ref{fullslicing2}), the shift, $\xi$, is
arbitrary, and
$\theta$ satisfies  $\theta = \frac{4}{\phi}\frac{\textrm{\scriptsize{d}}\phi}{\textrm{\scriptsize{d}}\lambda}$.
The $\phi$ variation gives the lapse-fixing equation (\ref{fullslicing}).  We emphasize that $\frac{\textrm{\scriptsize{d}}\phi}{\textrm{\scriptsize{d}}\lambda}$ is arbitrary in the full theories, unlike in the poor man's versions, where one ends up having to set the auxiliary ($\theta$ or $\nabla^2\eta$) to zero.  

It is not obvious then that $\dot{p} = 0$ 
is guaranteed form Hamilton's equations, since what one immediately obtains is, weakly,\fn{We use $\approx$ to 
denote weak
vanishing as defined by Dirac \cite{Dirac}, i.e equality modulo
constraints.}
\be
\begin{array}{l}
\frac{\pa p}{\pa \lambda} \approx  \frac{2N\sqrt{g}\phi^4}{V^{\frac{2}{3}}}\left(R\mbox{--}\frac{6\nabla^2\phi}{\phi}\right)- 
\frac{2\sqrt{g}}{\vc}  \nabla^2(\phi^4N)    + \frac{4\sqrt{g}}{\vc}g^{cd}(\phi^3N)_{,c}\phi_{,d} 
- \frac{2\sqrt{g}}{\vc}\phi^6\left<\phi^4N\left(R\mbox{--}\frac{8\nabla^2\phi}{\phi}\right)\right>.
\label{trham1}
\end{array}
\ee 
We now require use of  $\nabla^2(\phi^4N) = \phi\nabla^2(\phi^3N)+2g^{cd}(\phi^3N)_{,c}\phi_{,d}  
+ \phi^3N\nabla^2\phi$ to see that (\ref{fullslicing}) indeed guarantees $\dot{p} = 0$.  

A different application of the Dirac procedure used in this section is given for     
relativity in \cite{Niall-in-Dublin}.  One starts with the BSW
action (\ref{BSWac}), but omits the best-matching 3-diffeomorphism
corrections to the
bare velocities in the kinetic term. There is
no $\xi^{i}$ to vary in order to
obtain the momentum constraint.
However, if one again uses the evolution
equations to propagate
the quadratic square-root constraint, the momentum
constraint arises as a necessary integrability condition. This formalism is
of interest in establishing the extension of the 3-space approach
to include fermionic matter \cite{Vanderson}.
These issues in conformal gravity are beyond the scope of this paper and 
will be the subject of further work.

Finally, we briefly mention two more sources of variety in our family of theories.  
First, in RWR \cite{BOF}, the most general consistent BSW-type pure gravity action considered is 
\be I=\int\textrm{d}\lambda\sqrt{g}\sqrt{\Lambda+sR}
\sqrt{T_{\mbox{\scriptsize W \normalsize}}}.\ee  
For $s =1$,
$-1$ this gives $W = 1$ Lorentzian and Euclidean GR respectively, whilst for
$s = 0$ it gives \it strong gravity \normalfont \cite{stronggravity} generalized to arbitrary $W$ \cite{Sanderson}. 
We find that
there is analogously a \textit{strong conformal theory}
(which this time has without loss of generality $W = 0$), with action \be
    S^{\mbox{\scriptsize Strong \normalsize}} = \int \textrm{d}\lambda
    \frac{      \int   \textrm{d}^3x         \sqrt{g}\phi^6 \sqrt{\Lambda}
    \sqrt{    T^{\mbox{\scriptsize C\normalsize}}     }
    }{V},\label{ConStrongGrav}
\ee
which may be of use in understanding quantum conformal gravity.
Note that the power of the volume $V$ needed to make the
action homogeneous of degree zero is here one, since now it has to
balance only $\sqrt g$ and not the product $\sqrt{g}\sqrt{R}$.
This theory is simpler than conformal gravity in two ways:
$\Lambda$ is less intricate than $R$, and the lapse-fixing equation for strong conformal gravity is 
$\Lambda N = <\Lambda N>$ so since $<\Lambda N>$ is a spatial constant, $N$ is a spatial constant.

Second, instead of constructing an action with a local square
root, one could use instead a \textit{global} square root and thus
obtain 
\be 
S^{\mbox{\scriptsize Global\normalsize}} = \int
\textrm{d}\lambda \frac{\sqrt{ \int \textrm{d}^3x \sqrt{g}\phi^2\left(R - \frac{8\nabla^2\phi}{\phi}\right)}\sqrt{\int
\textrm{d}^3x\sqrt{g}\phi^6T^{\mbox{\scriptsize C\normalsize}}}}{\vc},  
\ee 
which gives rise to a single global
quadratic constraint.  
We note that the above alternatives are cumulative: for each of the theories in this section, we could consider 6 variants 
by picking Euclidean, strong, or Lorentzian signature and a local or global square root.  
The Lorentzian, local choices expanded above are the most obviously physical choice.  

\section{Integral Conditions and the Cosmological Constant}

We first demonstrate for pure conformal gravity that the
$\frac{2}{3}$ power that makes the action homogeneous is indeed
required.  For, if we change the action to \be
    S^{\mbox{\scriptsize $C_n$\normalsize}} = \int
    \textrm{d}\lambda \frac { \int \textrm{d}^3x\sqrt{g}\phi^4 \sqrt{ R -
    \frac{8\nabla^2\phi}{\phi} } \sqrt{T^{\mbox{\scriptsize
    C\normalsize}}  }    } {V^n }
\label{BOCCtrial} \ee with $n$ a free power, we get the modified
lapse-fixing equation (\ref{BMFLFC}) \be
    RN - \nabla^2N
    = \frac{3}{2}n<NR>
\label{arbslice} \ee in the distinguished representation.   Then 
integration of (\ref{arbslice}) over space yields
$$
    0 = \oint \sqrt{g}\textrm{d}S_a\nabla^aN =  \int
    \textrm{d}^3x\sqrt{g}\nabla^2N = \int \textrm{d}^3x\sqrt{g}RN - \frac{3}{2}n \int
    \textrm{d}^3x\sqrt{g(x)} \left( \frac{ \int \textrm{d}^3y\sqrt{  g(y)  }R(y)N(y) }{
    V  }\right)$$ \be = \left(1-\frac{3}{2}n\right)\int \textrm{d}^3x \sqrt{g}
    RN,\label{ConCheck}
\ee using, respectively, the fact that the manifold is compact
without boundary, Gauss's theorem, (\ref{arbslice}), and that
spatial integrals are constants and can therefore be pulled
outside further spatial integrals. This shows that consistency requires 
$n=2/3$,
since the integral cannot vanish. 
Likewise strong conformal gravity requires the power of the volume to be 1.

We recall that the above powers of the volume were found by
requiring the actions to be invariant under the constant rescaling
of $\phi$.  Thus, the integral consistency check (\ref{ConCheck})
indeed shows that this homogeneity invariance is indispensable for
the production of consistent actions in the compact case without
boundary.  If it is not observed, one obtains pathological frozen
dynamics: $N \equiv 0$.  We note that in contrast, in the
asymptotically-flat case considered in Sec. 4, we need not divide
by a volume term in the action, and the integral inconsistency 
argument is not applicable.

We now ask what happens when one attempts to combine the actions
of strong conformal gravity and conformal gravity in order to
consider conformal gravity with a cosmological constant. Applying
the homogeneity requirement, we obtain the combined action $$
    ^{\mbox{\scriptsize $\Lambda$\normalsize}}S = \int
    \textrm{d}\lambda \int \textrm{d}^3x \left( \left( \frac{\sqrt{g}\phi^6}{V}\right)
    \sqrt{\left(\frac{\vc}{\phi^4}\right)s\left( R -
    \frac{8\nabla^2\phi}{\phi}\right) + \Lambda }
    \sqrt{T^{\mbox{\scriptsize C\normalsize}}     }\right) $$ \be =
    \int \textrm{d}\lambda \frac{\int \textrm{d}^3x\sqrt{g}\phi^4 \sqrt{s \left( R -
    \frac{8\nabla^2\phi}{\phi} \right) + \frac{   \Lambda\phi^4   }{
    V(\phi)^{  \frac{2}{3}}   } } \sqrt{T^{\mbox{\scriptsize
    C\normalsize}}}} { V(\phi)^{\frac{2}{3}}  } = \int \textrm{d}\lambda
    \frac{\bar{J}}{V^{\frac{2}{3}}},
\label{Beast} \ee with cosmological constant $\Lambda$, where we
have also included a signature $s$ to show how (\ref{Beast})
reduces to the strong conformal gravity action
(\ref{ConStrongGrav}) in the limit $s \longrightarrow 0$.

The conjugate momenta $p^{ij}$ and $p_{\phi}$ are given by 
(\ref{confcanmom}) and (\ref{gravcanmom}) as before, but now with
$$
    2N = \sqrt{ \frac{
    T^{\mbox{\scriptsize C\normalsize}} } {   s \left( R - \frac{ 8
    \nabla^2 \phi }{ \phi } \right) + \frac{  \Lambda\phi^4 }{
    V(\phi)^{\frac{2}{3}}  }   } },
$$
and the primary constraint (\ref{primconstr}) holds. Again, the
end-point part of the $\phi$ variation yields $p_{\phi} = 0$, so
$p = 0$, so without loss of generality $W = 0$, but now the rest of
the $\phi$ variation gives a new lapse-fixing equation, \be
    2s(NR -
    \nabla^2N) + \frac{3N\Lambda}{\vc} = \frac{\bar{J}}{V} + \frac{<
    N\Lambda
    >}{\vc}
\label{CClapsefix} \ee in the distinguished representation. For
this choice of the action, there is indeed no integral
inconsistency: \be
    0 = 2s\oint \sqrt{g}\textrm{d}S_a\nabla^aN =  2s\int
    \textrm{d}^3x\sqrt{g}\nabla^2N = \int \textrm{d}^3x\sqrt{g}\left(2sRN + \frac{3N\Lambda}
{\vc} - 2N\left(sR + \frac{\Lambda}{\vc}\right) -
\frac{N\Lambda}{\vc}\right). \ee

The $\xi^{i}$-variation yields the usual momentum constraint
(\ref{mom}), and the local square root gives the constraint 
\be
    -^{\Lambda}{\cal H}^{\mbox{\scriptsize C\normalsize}}
    \equiv \frac{\sqrt{g}}{V^{\frac{2}{3}}}\left(sR +
    \frac{\Lambda}{V^{\frac{2}{3}}}\right) -
    \frac{\vc}{\sqrt{g}}g_{ik}g_{jl}p^{ij}p^{kl} = 0.
\ee 
in the distinguished representation.  
The Euler--Lagrange equations are \be
    \begin{array}{ll}
    \frac{\textrm{\scriptsize{d}} p^{ij}}{\textrm{\scriptsize{d}}\lambda} = &
    \frac{s\sqrt{g}N}{V^{\frac{2}{3}}}(g^{ij}R - R^{ij}) +
    \frac{\sqrt{g}s}{\vc}(\nabla^i\nabla^jN - g^{ij}\nabla^2N) -
    \frac{2N\vc}{\sqrt{g}}p^{im}{p^j}_m -
    \frac{\bar{J}\sqrt{g}}{V^{\frac{5}{3}}}g^{ij} +
    \frac{\Lambda\sqrt{g}}{V^{\frac{4}{3}}}g^{ij}\left( N -
    \frac{<N>}{3}\right),
\label{LambdaEL}
\end{array}
\ee where we have split the working up into pure conformal gravity
and $\Lambda$ parts (the $\bar{J}$ here can also be split into the
pure conformal gravity integrand and a $\Lambda$ part).

Then by (\ref{CClapsefix}) $\dot{p} \approx 0$,  and finally 
\be
    -^{\Lambda}\dot{\cal H}^{\mbox{\scriptsize C\normalsize}}
\approx \frac{s\sqrt{g}}{V^{\frac{2}{3}}}\frac{\pa R}{\pa
    \lambda} - \frac{2\vc}{\sqrt{g}}\left( \frac{\pa p^{ij}}{\pa\lambda}
    p_{ij} + \frac{\pa g_{ik}}{\pa\lambda} p^{ij}{p^k}_j\right)
    \approx 0
\ee 
by the use of (\ref{nog}) and (\ref{nov}) in the first step,
and in the second step $\frac{\pa p^{ij}}{\pa\lambda} = \frac{\pa
    p^{ij}}{\pa\lambda}|_{\Lambda\mbox{\scriptsize -free\normalsize}} + 
\frac{\pa
    p^{ij}}{\pa\lambda}|_{\Lambda}$
and $\frac{\pa
    p^{ij}}{\pa\lambda}|_{\Lambda} \propto g^{ij}$,
so $p_{ij}$ annihilates this term, thus reducing
$^{\Lambda}\dot{{\cal H}}^{\mbox{\scriptsize C\normalsize}}$ to
the pure conformal gravity  $\dot{{\cal H}}^{\mbox{\scriptsize
C\normalsize}} \approx 0$.

Note that in conformal gravity the cosmological constant
$\Lambda$ (just like its particle model analogue, the Newtonian
energy $E$) contributes to a conformally-induced cosmological-constant
type force. The penultimate term in (\ref{LambdaEL}) is the final term
of pure conformal gravity in (\ref{BMFGravEL}), and the final term is
induced by $\Lambda$. Next, we will see
that matter also gives analogous contributions. The significance
of this is discussed in Sec. 8.

\section{Conformal Gravity Coupled to Matter Fields: General Results}

It is an important test of our theoretical framework to see
whether it is capable of accommodating enough classical field
theories to be a viable description of nature.  In RWR \cite{BOF,
AB}, we showed that if matter fields are `added on', GR imposes a
universal light cone and requires 3-vector fields to be gauge
fields.  We now show below how these results carry over to
conformal gravity.

In conformal gravity, we have to check the propagation of three
different constraints. The momentum constraint ${\cal H}_i$ is never
problematic, but the other two -- the quadratic constraint 
${\cal H}^{\mbox{\scriptsize C\normalsize}}$ that arises from the local
square-root form of the BSW action and the linear conformal
constraint $p=0$ -- need to be carefully checked.  The momentum and quadratic square-root constraints were shown in RWR \cite{BOF} 
to create the Efe's, the universal light cone obeyed by the bosonic matter and gauge  
theory. This paper shows how the matter results also hold in conformal gravity where the linear conformal constraint is also present.  

We begin with two theorems that suffice for the construction of a
range of classical field theories coupled to conformal gravity.
The first is about homogeneity and the propagation of $p = 0$ and
the second is about the propagation of ${\cal
H}^{\mbox{\scriptsize C\normalsize}}$. We will then demonstrate
example by example that the range of theories covered by these
theorems includes all of known classical bosonic physics coupled
to conformal gravity.  Furthermore, these theories are singled out
from more general possibilities by exhaustive implementation of
Dirac's demand for dynamical consistency.

Let $\Psi$ be a set of matter fields that we wish to couple to
conformal gravity, with potential term $^{\Psi}U$ and kinetic term
$^{\Psi}T$. We first decompose these as polynomials in the inverse
metric. This is because it is the power of the metric that
determines the powers of $V$ that must be used to achieve the
necessary homogeneity. Let $\Phi^{(n)}$ be the set of fields such
that these polynomials are of no higher degree than n. Thus

\noindent$ \mbox{ }^{\Psi^{(n)}}T = \Sigma_{(k =
0)}^{(n)}T^{(k)}_{i_1j_1i_2j_2 ... i_kj_k}g^{i_1j_1} ... g^{i_kj_k} =
\Sigma_{(k = 0)}^{(n)}T_{(k)}$, $  ^{\Psi^{(n)}}U = \Sigma_{(k =
0)}^{(n)}U^{(k)}_{i_1j_1i_2j_2 ... i_kj_k}g^{i_1j_1} ... g^{i_kj_k}  =
\Sigma_{(k = 0)}^{(n)} U_{(k)}$.

Then the following theorem guarantees that $p = 0$ is preserved by
the dynamical evolution.

\mbox{ }

\mbox{ }

\noindent\bf Theorem 1 \normalfont

\noindent For matter fields $\Psi^{(n)}$, the conformal gravity +
matter action of the form \be
    ^{\Psi^{(n)}}S = \int \textrm{d}\lambda\frac {      \int
    \textrm{d}^3x\sqrt{g}\phi^4 \sqrt{ s\left( R - \frac{8\nabla^2\phi}{\phi}
    \right) + \frac{ \phi^4 }{ V^{\frac{2}{3}} }\Sigma_{(k =
    0)}^{(n)}\frac{  U_{(k)} V^{\frac{2k}{3}}  }{  \phi^{4k}  }   }
    \sqrt{ T^{\mbox{\scriptsize C\normalsize}} + \Sigma_{(k = 0)}^{(n)}
\frac{  T_{(k)}V^{\frac{ 2k }{
    3 }}  }{ \phi^{4k}  }   } } {     V^{\frac{2}{3}}     }
\ee varied with free end points is guaranteed to have $\dot{p} =
0$ $\forall$ $n \in {\cal N}_0$.

Note how the powers of $V$ match
the powers of the inverse metric that are needed to make
3-diffeomorphism scalars from the matter fields of different
possible ranks.

\noindent\bf Proof \normalfont Vacuum conformal gravity holds,
hence the theorem is true for $n = 0$.

\noindent Induction hypothesis: suppose the theorem is true for
some $n = q$.

\noindent Then, for $n = q + 1$, $\phi$ variation gives \be
    0 =
    \frac{\delta S^{(q + 1)}}{ \delta\phi(x)} = \frac{\delta
    S^{(q)}}{\delta\phi(x)}
\begin{array}{l}
    + 4(2 - q)NV^{\frac{2(q - 1)}{3}}U_{(q + 1)} - \frac{q + 1}{N}T_{(q + 
1)}V^{\frac{2}{3q}} \\
    + 4\int \textrm{d}^3x\sqrt{g}V^{\frac{2q - 5}{3}} \left( qNU_{(q + 1)} +
    \frac{q + 1}{4N}V^{\frac{2}{3}}T_{(q + 1)} \right).
\end{array}
\label{line1} \ee
Now, from $\dot{p}^{(q + 1)} = \dot{p}^{(q + 1)ij}g_{ij} + p^{(q +
1)ij}\dot{g}_{ij}$ and the metric Euler--Lagrange
equation for $\dot{p}^{(q + 1)ij}$, 
\be
\begin{array}{ll}
    \dot{p}^{(q + 1)} = & \dot{p}^{(q)} + \int \textrm{d}^3x\sqrt{g}V^{\frac{2q
    -5}{3}} \left( NU_{(q + 1)} + \frac{   V^{  \frac{2}{3}  }   }{ 4N
    }T_{(q + 1)} \right) + 3N\sqrt{g}U_{(q + 1)}V^{  \frac{2(q -
    2)}{3}  } \\ & + V^{  \frac{2(q - 2)}{3}  }\sqrt{g} \left( \frac{
    V^{ \frac{2}{3} }   }{   4N   } \frac{   \delta T_{(q + 1)}   }{
    \delta g_{ij}   } + N\frac{   \delta U_{(q + 1)}   }{   \delta
    g_{ij}   } \right)g^{ij}
\end{array}
\ee Hence, by (\ref{line1}) \be
    \dot{p}^{(q + 1)} =  V^{ \frac{2(q
    -2)}{3} }\sqrt{g} \left( \frac{  V^{ \frac{2}{3} }  }{  4N  }
    \left( \frac{\delta T_{(q + 1)}}{\delta g_{ij}}g^{ij} + (q +
    1)T_{(q + 1)} \right) + N \left( \frac{\delta U_{(q + 1)}}{\delta
    g_{ij}}g^{ij} + (q + 1)U_{(q + 1)} \right) \right)
    = 0
\ee by the induction hypothesis and using that $U_{(q + 1)}$, $T_{(q +
1)}$ are homogeneous of degree $q + 1$ in $g^{ij}$. Hence, if the
theorem is true for $n = q$, it is also true for $n = q + 1$.  But
it is true for $n = 0$, so it is true by induction $\forall$ $n
\in {\cal N}_0$. $\Box$

We will now consider $^{\Psi}T$ and $^{\Psi}U$ as being made up of
contributions from each of the fields present.  We will label
these fields, and the indices they carry, by capital Greek
indexing sets.  We then obtain the following formulae for the
propagation of the Hamiltonian constraint.

\noindent \bf Theorem 2 \normalfont

\noindent i) For nonderivative coupled matter fields
$\Psi_{\Delta}$ with $^{\Psi}T$ homogeneously quadratic in
$\dot{\Psi}_{\Delta}$ and $^{\Psi}U$ containing at most
first-order derivatives, \be
    -^{\Psi}\dot{{\cal H}}^{\mbox{\scriptsize C\normalsize}} =
    \frac{1}{N}\nabla_b \left(
    N^2\left(2G_{\Delta\Gamma}\Pi^{\Gamma}\frac{ \pa ^{\Psi}U }{
    \pa(\nabla_b\Psi_{\Delta})  } + s\left[\Pi^{\Gamma}\frac{\delta
    (\pounds_\xi\Psi_{\Gamma}) }{\pa\xi_b} \right] \right)\right),
\ee where [ ] denotes the extent of applicability of the
functional derivative within, $G_{\Gamma\Delta}$ is an invertible
ultralocal kinetic metric and $\Pi^{\Gamma}$ is the
momentum conjugate to $\Psi_{\Delta}$.

\noindent ii) If, additionally, the potential contains covariant
derivatives, then there is an extra contribution to i): \be
    \frac{2\sqrt{g}}{N}\nabla_b\left( N^2p_{ij}\left(\frac{\pa
    ^{\Psi}U}{\pa{\Gamma^a}_{ic}}g^{aj} - \frac{1}{2}\frac{\pa
    ^{\Psi}U}{\pa{\Gamma^a}_{ij}}g^{ac}\right)\right).
\ee

The proofs offered here include both conformal gravity ($s = 1$)
and strong conformal gravity ($s = 0$, $\Lambda \neq 0$).  There
is an equivalent derivation which encompasses GR and the
arbitrary-$W$ strong gravity theories.  Result i) in the GR case is
related to a result of Teitelboim \cite{Teitelboim} that
the contributions of nonderivatively-coupled fields to the Hamiltonian
and momentum constraints independently satisfy the Dirac algebra.
In the working below, this is reflected by our ability to split the
working into pure gravity and matter parts.

Use of formulae i), ii) permits the $^{\Psi}\dot{{\cal
H}}^{\mbox{\scriptsize C\normalsize}}$ calculations to be done
without explicitly computing the Euler--Lagrange equations. This
is because our derivation uses once and for all the \it general
\normalfont Euler--Lagrange equations.

\noindent \bf Proof \normalfont

\noindent i) For a homogeneous quadratic kinetic term $
    ^{\Psi}T = (\dot{\Psi}_{\Gamma} -
    \pounds_{\xi}\Psi_{\Gamma})(\dot{\Psi}_{\Delta} -
    \pounds_{\xi}\Psi_{\Delta})G^{\Gamma\Delta}\mbox{\scriptsize
    $\left(\frac{\vc}{\phi^4}g^{ij}\right)$\normalsize},$

    \noindent the conjugate momenta are
    $\Pi^{\Delta} = \frac{\pa L
    }{\pa\dot{\Psi}_{\Delta}} = \frac{\sqrt{g}\phi^4}{2N\vc}{
    }G^{\Gamma\Delta}(\dot{\Psi}_{\Gamma} -
    \pounds_{\xi}\Psi_{\Gamma})$.

    \noindent The $\xi^i$-variation gives the momentum constraint
    \be -^{\Psi}{\cal H}^{\mbox{\scriptsize C\normalsize}}_i \equiv 
2\nabla_j{p_i}^j
    -
    \Pi^{\Delta}\frac{\delta(\pounds_{\xi}\Psi_{\Delta})}{\delta\xi^i}
    = 0 \label{genconfmom}
\ee and the local square root gives the Hamiltonian constraint,
which is 
\be
    -^{\Psi}{\cal H}^{\mbox{\scriptsize C\normalsize}} \equiv 
\frac{  \sqrt{g}  }{   V^{  \frac{2}{3}  }   }(sR + {}^{\Psi}U)  
- \frac{ \vc  }{  \sqrt{g}  }(p^{ij}p_{ij} + G_{\Delta\Gamma}\Pi^{\Gamma}\Pi^{\Delta}) = 0
\label{genconfham} 
\ee 
in the distinguished representation. Then
\be
\begin{array}{ll}
    -^{\Psi}\dot{{\cal H}}^{\mbox{\scriptsize C\normalsize}} \approx
    \frac{\sqrt{g}}{V^{\frac{2}{3}}}(s\dot{R} + {}^{\Psi}\dot{U}) -
    \frac{2\vc}{\sqrt{g}}(\dot{p}^{ij}p_{ij} + \dot{g}_{ik}p^{ij}{p^k}_j)
    - \frac{\vc}{\sqrt{g}}(2\dot{\Pi}^{\Delta} {
    }G_{\Gamma\Delta}\Pi^{\Gamma} +  {
    }\dot{G}_{\Gamma\Delta}\Pi^{\Delta}\Pi^{\Gamma}),
\end{array}
\ee 
using the chain-rule on (\ref{genconfham}) and using $\dot{g}
= \dot{V} = 0$.  Now use the chain-rule on $^{\Psi}\dot{U}$, the
Euler--Lagrange equations $\dot{p}^{ij} = \frac{\delta L}{\delta
g_{ij}}$ and $\dot{\Pi}^{\Delta} = \frac{\delta L}{\delta
\Psi_{\Delta}}$, and $p = 0$ to obtain the first step below: 
\be
\begin{array}{ll}
-^{\Psi}\dot{{\cal H}}^{\mbox{\scriptsize C\normalsize}} \approx &
\frac{\sqrt{g}s}{V^{\frac{2}{3}}}\dot{R}
+ \frac{\sqrt{g}}{V^{\frac{2}{3}}}\left(
\frac{\pa{}^{\Psi}U}{\pa\Psi_{\Delta}}\dot{\Psi}_{\Delta} 
+ \frac{\pa{}^{\Psi}U}{\pa(\nabla_b\Psi_{\Delta})}\dot{(\nabla_b\Psi_{\Delta})}
+ \frac{\pa{}^{\Psi}U}{\pa g_{ab}}\dot{g}_{ab}\right)
\\ &
- {2s}p^{ij}\left[\frac{\delta R}{\delta{g_{ij}}}N\right] 
- {2}p^{ij}\left[\frac{\delta^{\Psi}U}{\delta{g_{ij}}} N\right] 
- \frac{1}{2N}p^{ij}\frac{\pa ^{\Psi}T}{\pa{g_{ij}}} 
- \frac{4N\vc}{{g}}p_{ik}p^{ij}{p^k}_j
\\ &
- {2}{ }G_{\Gamma\Delta}\Pi^{\Gamma}\left[N\frac{\delta U_{\Psi}}{\delta\psi_{\Delta}} \right] 
- \frac{\vc}{\sqrt{g}}\mbox{}\dot{G}_{\Delta\Gamma}\Pi^{\Delta}\Pi^{\Gamma}
\\

\mbox{  }\mbox{  }\mbox{  }\mbox{  }\mbox{  } \mbox{  } \mbox{  }
= & \left(\frac{\sqrt{g}s}{V^{\frac{2}{3}}}\dot{R} 
- {2s}p^{ij}\left[\frac{\delta R}{\delta{g_{ij}}} N\right] 
- \frac{4N\vc}{{g}}p_{ik}p^{ij}{p^k}_j\right) \\ &

+\frac{\sqrt{g}}{V^{\frac{2}{3}}}\left(\frac{\pa
{}^{\Psi}U}{\pa\Psi_{\Delta}}\left(\frac{2N\vc}{\sqrt{g}}\Pi^{\Gamma}G_{\Gamma\Delta}\right)

+ \frac{\pa
{}^{\Psi}U}{\pa(\nabla_b\Psi_{\Delta})}\nabla_b\left(\frac{2N\vc}{\sqrt{g}}\Pi^{\Gamma}
G_{\Gamma\Delta}\right) + \frac{\pa {}^{\Psi}U}{\pa g_{ab}}\left(
\frac{2N\vc}{\sqrt{g}}p_{ab} \right)\right) \\ &

- {2}p^{ij}\frac{\pa {}^{\Psi}U}{\pa{g_{ij}}} N 
- \frac{1}{2N}p^{ab}\frac{\pa{ }G^{\Delta\Gamma}}{\pa
g_{ab}}\dot{\Psi}_{\Gamma}\dot{\Psi}_{\Delta} \\&

- {2}{ }G_{\Gamma\Delta}\Pi^{\Gamma} N\frac{\pa{}^{\Psi}U}{\pa\psi_{\Delta}}

+ {2}G_{\Gamma\Delta}\Pi^{\Gamma}\nabla_b\left(N\frac{\pa{}^{\Psi}U} {\pa(\nabla_b\Psi_{\Delta})}\right)

- \frac{\vc}{\sqrt{g}}\frac{\pa{G}_{\Delta\Gamma}}{\pa
g_{ij}}\dot{g}_{ij}\Pi^{\Delta}\Pi^{\Gamma}

\\

\mbox{  }\mbox{  } \mbox{  } \mbox{  }\mbox{ }\mbox{ }\mbox{ }
\approx & \frac{s}{N}\nabla_b\left(N^2\Pi^{\Delta}\frac{\delta(\pounds_{\xi}\Psi_{\Delta})}{\delta\xi_b}\right)
+ \sqrt{g}\frac{\pa{}^{\Psi}U}{\pa(\nabla_b\Psi_{\Delta})}\nabla_b\left(\frac{2N}{\sqrt{g}}\Pi^{\Gamma}{ }G_{\Delta\Gamma}\right) 
+ {2}G_{\Gamma\Delta}\Pi^{\Gamma}\nabla_b\left(N\frac{\pa{}^{\Psi}U} {\pa(\nabla_b\Psi_{\Delta})}\right).
\end{array}
\ee 
In the second step above, we regroup the terms into pure
gravity terms and matter terms, expand the matter variational
derivatives and use the definitions of the momenta to eliminate
the velocities in the first three matter terms.  We now observe
that the first and sixth matter terms cancel, as do the third and
fourth.  In the third step we use the pure gravity working and the
momentum constraint (\ref{genconfmom}), and the definitions of the
momenta to cancel the fifth and eight terms of step 2.
Factorization of step 3 gives the result.

ii) Now $-^{\Psi}\dot{{\cal H}}^{\mbox{\scriptsize C\normalsize}}$
has 2 additional contributions in step 2 due to the presence of
the connections: \be
    \frac{\sqrt{g}}{V^{\frac{2}{3}}}\frac{\pa
    U_{\Psi}}{\pa{\Gamma^a}_{bc}}  \dot{\Gamma}^a {}_{bc} -
    {2}p^{ij}\left[\frac{\pa
    U_{\Psi}}{\pa{\Gamma^a}_{bc}}\frac{\delta {\Gamma^a}_{bc}}{\delta
    g_{ij}}N\right],
\label{halfwaycon} \ee which, using \be
    \delta{\Gamma}^a {}_{bc} = \frac{1}{2}g^{ad}(\nabla_c(\delta
    g_{db}) + \nabla_b(\delta g_{dc})- \nabla_d(\delta g_{bc})),\ee \be
    \dot{\Gamma}^a {}_{bc} = \frac{\vc}{2\sqrt{g}}(\nabla_b(N{p_c}^a) +
    \nabla_c(N{p_b}^a) - \nabla^a(Np_{bc})),
\ee integration by parts on the second term of (\ref{halfwaycon})
and factorization yields ii). $\Box$

Although Theorem 1 does not consider potentials containing
Christoffel symbols, in  all the cases that we consider below
(which suffice for the investigation of the classical bosonic
theories of nature) the propagation of $^{\Psi}{\cal
H}^{\mbox{\scriptsize C\normalsize}}$ rules out all theories
with such potentials. Thus it is not an issue whether such
theories permit $p = 0$ to be propagated.

\section{Conformal Gravity Coupled to Matter Fields: Examples}

In this section we take $s = 1$ for Lorentzian (as opposed to
Euclidean or strong) conformal gravity.  We will also use $W = 0$
from the outset, and $\Lambda = 0$, so that we are investigating
whether our theory of pure conformal gravity is capable of
accommodating conventional classical matter theories and
establishing the physical consequences. We find that it does, and
that the known classical bosonic theories are singled out.

\subsection{ Scalar Field }

The natural action to consider according to our prescription for
including a scalar field is \be ^{\psi}S_{\mbox{\scriptsize
\normalsize}} =\int \textrm{d}\lambda \int \textrm{d}^3x\left(
\left(\frac{\sqrt{g}\phi^6}{V}\right)
\sqrt{\left(\frac{\vc}{\phi^4}\right)\left( R -
\frac{8\nabla^2\phi}{\phi} + {}^{\psi}U_{(1)} \right) +
{}^{\psi}U_{(0)} } \sqrt{T^{\mbox{\scriptsize C\normalsize}} +
{}^{\psi}T    }\right) \ee \be = \int d\lambda \frac{\int
\textrm{d}^3x\sqrt{g}\phi^4 \sqrt{ R - \frac{8\nabla^2\phi}{\phi}  +
{}^{\psi}U_{(1)} + \frac{ {}^{\psi}U_{(0)}\phi^4 }{   V(\phi)^{
\frac{2}{3}}   } } \sqrt{T^{\mbox{\scriptsize C\normalsize}} +
{}^{\mbox{\scriptsize $\psi$\normalsize}}T}} {
V(\phi)^{\frac{2}{3}}  } = \int \textrm{d}\lambda
\frac{\bar{I}}{V^{\frac{2}{3}}}, \label{BOSaction} \ee where, once
again, we give two different expressions to exhibit the
homogeneity and to use in calculations. $^{\psi}U_{(0)}$
is an arbitrary function of $\psi$ alone whilst $^{\psi}U_{(1)} =
-\frac{C}{4}g^{ab}(\pa_a\psi)\pa_b\psi$. 
$\sqrt{C}$ is the a priori unfixed canonical speed of propagation of the scalar field.

The conjugate momenta $p^{ij}$ and $p_{\phi}$ are given by
(\ref{gravcanmom}) and (\ref{confcanmom}) but with
$$
    2N = \sqrt{
    \frac{ T^{\mbox{\scriptsize C\normalsize}} + {}^{\mbox{\scriptsize
    $\psi$\normalsize}}T } {    R - \frac{ 8 \nabla^2 \phi }{ \phi }  +
    {}^{\psi}U_{(1)} + \frac{{}^{\psi}U_{(0)}\phi^4 }{ V(\phi)^{\frac{2}{3}} 
} } },
$$
and additionally we have the momentum conjugate to $\psi$, $
    p_{\psi} = \frac{\sqrt{g}\phi^4}{2N}(\dot{\psi} -
    \pounds_{\xi}\psi)$.
As in the case of pure conformal gravity, we have the primary
constraint (\ref{primconstr}), and the end-point part of the
$\phi$-variation gives $p_{\phi} = 0$, so that $p = 0$ by the
primary constraint. But by construction (Theorem 1) this action
has the correct form to propagate the constraint $p = 0$ provided
the lapse-fixing equation 

\noindent\be
    2(NR -\nabla^2N) + \frac{3N^{\psi}U_{(0)}}{\vc} + 2N{}^{\psi}U_{(1)} =
    \frac{1}{V^{\frac{5}{3}}}\int \textrm{d}^3x\sqrt{g}N{}^{\psi}U_{(0)} +
    \frac{\bar{I}}{V},
\ee holds (in the distinguished representation), but this is guaranteed from the
rest of the $\phi$-variation.

\noindent The $\xi^i$-variation gives the momentum constraint $
    -^{\psi}{\cal
    H}^{\mbox{\scriptsize C\normalsize}}_i \equiv 2\nabla_j{p_i}^j - 
p_{\psi}\pa_i\psi = 0$,

\noindent whilst the local square root gives rise to the
Hamiltonian constraint, which is 
\be
    -^{\psi}{\cal H}^{\mbox{\scriptsize C\normalsize}} \equiv
    \frac{\sqrt{g}}{V^{\frac{2}{3}}} \left(R + {}^{\psi}U_{(1)} +
    \frac{^{\psi}U_{(0)}}{\vc}\right) - \frac{\vc}{\sqrt{g}}(p^{ij}p_{ij} +
    p_{\psi}^2) = 0
\ee 
in the distinguished representation.

Then, using formula i), the propagation of the Hamiltonian
constraint is, weakly, 
\be
^{\psi}\dot{\cal H}^{\mbox{\scriptsize C\normalsize}} \approx
\frac{(C - 1)}{N}\nabla_b(N^2p_{\psi}\pa^b\psi).
\ee
Now, if the cofactor of $(C - 1)$ were zero, there would be a
secondary constraint which would render the scalar field theory
trivial by using up its degree of freedom. Hence $C = 1$ is fixed, which is the universal light-cone condition applied to the scalar field.  
This means that the same light cone as that of the gravitation is enforced even though the
gravitational theory in question is not generally covariant.  

\subsection{1-Form Fields}

According to our prescription, the natural action to include
electromagnetism is \be ^{\mbox{\scriptsize A\normalsize}}S = \int
\textrm{d}\lambda \int \textrm{d}^3x\left( \left( \frac{  \sqrt{g}\phi^6  }{  V  }
\right) \sqrt{ \left( \frac{  \vc  }{  \phi^4  } \right) \left( R
- \frac{ 8\nabla^2\phi  }{  \phi  } \right) + \left( \frac{    V^{
\frac{4}{3}  }}{  \phi^8  } \right) {}^{\mbox{\scriptsize
A\normalsize}}U   } \sqrt{ T^{\mbox{\scriptsize C\normalsize}} +
\left( \frac{ \vc }{ \phi^4 } \right) {}^{\mbox{\scriptsize
A\normalsize}}T } \right) \ee \be = \int \textrm{d}\lambda \frac{ \int
\textrm{d}^3x\sqrt{g}\phi^4 \sqrt{        R - \frac{8\nabla^2\phi}{\phi} +
{}^{\mbox{\scriptsize A\normalsize}}U\frac{   V(\phi)^{
\frac{2}{3} } } {    \phi^4    }         } \sqrt{
T^{\mbox{\scriptsize C\normalsize}} + {}^{\mbox{\scriptsize
A\normalsize}}T\frac{ V(\phi)^{ \frac{2}{3} } } {  \phi^4  }
}          } { V(\phi)^{ \frac{2}{3} } } = \int \textrm{d}\lambda \frac{
\bar{I} }{V^{\frac{2}{3}}}, \label{BOCCaction} \ee for \be
    ^{\mbox{\scriptsize A\normalsize}}U =
    -\hat{C}(\nabla_bA_a - \nabla_aA_b)\nabla^bA^{a}, \ee \be
    ^{\mbox{\scriptsize A\normalsize}}T = g^{ab}(\dot{A}_a -
    \pounds_{\xi}A_a)(\dot{A}_b - \pounds_{\xi}A_b).
\label{emT} \ee We will first show that electromagnetism exists as
a theory coupled to conformal gravity. We will then discuss how it
is uniquely picked out (much as it is picked out in RWR
\cite{BOF}), and how Yang--Mills theory is uniquely picked out
upon consideration of K interacting 1-form fields (much as it is
picked out in \cite{AB}).  

Again, the conjugate momenta $p_{\phi}$ and $p^{ij}$ are given by
(\ref{confcanmom}) and (\ref{gravcanmom}) but now with
$$
    2N = \sqrt{
    \frac{T^{\mbox{\scriptsize C\normalsize}} +
    \frac{{}^{\mbox{\scriptsize A\normalsize}}T V(\phi)^{\frac{2}{3}}
    }{\phi^4 } } {   s \left( R - \frac{ 8 \nabla^2 \phi }{ \phi }
    \right) + \frac{{}^{\mbox{\scriptsize A\normalsize}}U
    V(\phi)^{\frac{2}{3}} }{\phi^4 } } },
$$
and additionally we have the momentum conjugate to $A_i$, $
    \pi^i =
    \frac{\sqrt{g}}{2N}g^{ij}(\dot{A}_j - \pounds_{\xi}A_j)$.

\noindent By the same argument as in previous sections, $p = 0$
arises and is preserved by a lapse-fixing equation, which is now 

\noindent\be
    2(NR
    -\nabla^2N) +\left( N{}^{\mbox{\scriptsize A\normalsize}}U -
    \frac{{}^{\mbox{\scriptsize A\normalsize}}T}{4N} \right)\vc +
    \frac{1}{V^{\frac{1}{3}}}\int
    \textrm{d}^3x\sqrt{g}\left(N{}^{\mbox{\scriptsize A\normalsize}}U +
    \left(\frac{{}^{\mbox{\scriptsize A\normalsize}}T}{4N}\right)\right)
    = \frac{\bar{I}}{V} \label{VECTORSLICE}
\ee The $\xi^i$-variation gives the momentum constraint
$-^{\mbox{\scriptsize A\normalsize}}{\cal
    H}^{\mbox{\scriptsize C\normalsize}}_i \equiv 2\nabla_j{p_i}^j - 
(\pi^{c}(\nabla_i{A_{c}} -
    \nabla_c{{A}_i}) - {\nabla_c\pi^{c}}A_{i}) = 0$,
whilst the local square root gives rise to the Hamiltonian
constraint, which is 
\be
    -^{\mbox{\scriptsize A\normalsize}}{\cal H}^{\mbox{\scriptsize C\normalsize}} \equiv 
\frac{\sqrt{g}}{V^{\frac{2}{3}}} \left(sR 
+ {{}^{\mbox{\scriptsize A\normalsize}}U}{\vc}\right) 
- \frac{\vc}{\sqrt{g}}(p^{ij}p_{ij} + \frac{1}{\vc}\pi_i\pi^i) = 0
\ee 
in the distinguished representation.

Then, using formula i), the propagation of the Hamiltonian
constraint is, weakly, 
\be
    -^{\mbox{\scriptsize A\normalsize}}\dot{\cal H}^{\mbox{\scriptsize 
C\normalsize}}
\approx \frac{1}{N} \nabla^b \left( N^2 \left( (1
    - 4\hat{C})\pi^i(\nabla_bA_i - \nabla_iA_b) - A_b\nabla_i\pi^i \right)
    \right)
\ee Suppose the cofactor of $1 - 4\hat{C}$ is zero. Then we require
$\nabla_{[b}A_{i]} = 0$.  But this is three conditions on $A_i$, so
the vector theory would be rendered trivial. Thus, exhaustively,
the only way to obtain a consistent theory is to have the universal light-cone condition 
$\hat{C} = 1/4$ and the new constraint \be {\cal G} \equiv \nabla_a\pi^a = 0, \ee
which is the electromagnetic Gauss constraint. The propagation of
${\cal G}$ is no further bother because the $A_i$ Euler--Lagrange
equation \be
    \frac{ \textrm{d}\pi^i }{ \textrm{d}\lambda} =
    2\sqrt{g}C\nabla_b(\nabla^bA^i - \nabla^iA^b)
\label{emEL} \ee is free of V and hence identical to that in the
RWR case. Since the RWR argument for the propagation of ${\cal G}$
follows from (\ref{emEL}), this guarantees that the result also
holds in conformal gravity.

We can finally encode this new constraint by making use of the best
matching associated with the U(1) symmetry of the potential,
to modify the kinetic term (\ref{emT}) by the introduction of an
auxiliary variable $\Phi$ to \be
    ^{\mbox{\scriptsize A\normalsize}}T =
    (\dot{A}_a - \pounds_{\xi}A_a - \pa_a\Phi)(\dot{A}^a -
    \pounds_{\xi}A^a - \pa^a\Phi).
\ee

The following extensions of this working have been considered.

1) Additionally, replacing $^{\mbox{\scriptsize A \normalsize}}U$
by $C^{abcd}\nabla_bA_a \nabla_dA_c$ for $C^{abcd} =
C_1g^{ac}g^{bd} + C_2g^{ad}g^{bc} + C_3g^{ab}g^{cd}$ in the action
preserves the correct form to guarantee $p = 0$ is maintained. We
now have derivative coupling contributions also, so we need to
make use of formula ii) of theorem 2 as well as formula i). Thus,
weakly \be
    -^{\mbox{\scriptsize A\normalsize}}\dot{{\cal H}}^{\mbox{\scriptsize 
C\normalsize}}
\approx \frac{1}{N}\nabla_b \left(
    N^2 \left(
    \begin{array}{l}
    (4C_1 + 1)\pi^a\nabla^b{A_a} + (4C_2 - 1)\pi^a\nabla_a{A^b} +
    4C_3(N^2\pi^b\nabla_a{A^a}) \\ - \nabla_a{\pi^a}A^b -
    4p_{ij}\nabla_{(d}A_{b)}\left(C^{ajbd}A^i - \frac{1}{2}C^{ijbd}A^a
    \right).
    \end{array}
    \right) \right)
\ee This has the same structure in $A_i$ as for the GR case [the
overall $V^{-\frac{2}{3}}$ is unimportant, as is the replacement of
the GR $(p_{ij} - \frac{p}{2}g_{ij})$ factors by $(p_{ij})$
factors here], so an argument along the same lines as that used in
RWR will hold, forcing the Gauss constraint and $C_1 = - C_2 = -\frac{1}{4}$,
$C_3 = 0$ (Maxwell theory).

2) The changes \be
    {}^{\mbox{\scriptsize A\normalsize}}T \longrightarrow \
{}^{{\mbox{\scriptsize A\normalsize}}_I}T =
    g^{ij}(\dot{A}_i^I - \pounds_{\xi}A_i^I) (\dot{A}_{jI} -
    \pounds_{\xi}A_{jI}),\label{TAISUB} \ee \be {}^{\mbox{\scriptsize 
A\normalsize}}
U \longrightarrow
    {}^{\mbox{\scriptsize A\normalsize}_I}U =  O_{IK}C
    ^{abcd}\nabla_bA^{I}_{a}\nabla_dA^K_{c} + {B^I}_{JK}\bar{C
    }^{abcd}\nabla_bA_{Ia}A^J_c A^K_d + I_{JKLM}\bar{\bar{C
    }}^{abcd}A^J_aA^K_bA^L_cA^M_d
\label{UAISUB} \ee (for a priori distinct supermetrics $C$,
$\bar{C}$, $\bar{\bar{C}}$) to the ansatz preserve the conformal
properties, hence guaranteeing that $p = 0$ is maintained by the
lapse-fixing equation obtained by applying (\ref{TAISUB},
\ref{UAISUB}) to (\ref{VECTORSLICE}).  The new conjugate momenta
are $
    \pi^i_I =
    \frac{\sqrt{g}}{2N}g^{ij}(\dot{A}_{Ij} -
    \pounds_{\xi}A_{Ij})$.

\noindent The $\xi^i$-variation gives the momentum constraint, \be
   -^{\mbox{\scriptsize A\normalsize}_I}{\cal H}^{\mbox{\scriptsize 
C\normalsize}}_i
    = 2\nabla_j{{p^j}_i} - (\pi^{Ij}(\nabla_i{A_{Ij}} -
    \nabla_j{{A_I}_i}) - \nabla_j{\pi_I^j}A^I_i) = 0,\ee
and the local square root gives the Hamiltonian constraint, which is \be
   - ^{\mbox{\scriptsize A\normalsize}_I}{\cal H}^{\mbox{\scriptsize 
C\normalsize}} \equiv
    \frac{\sqrt{g}}{V^{\frac{2}{3}}} \left(sR + {U_{\mbox{\scriptsize
    A\normalsize}_I}}{\vc}\right) - \frac{\vc}{\sqrt{g}}(p^{ij}p_{ij} +
    \frac{1}{\vc}\pi^I_i\pi_I^i) = 0
\ee in the distinguished representation.

Using formulae i), ii) we read off that the propagation of the
Hamiltonian constraint is, weakly, \be
    -{}^{{\mbox{\scriptsize A\normalsize}}_I}\dot{{\cal 
H}}^{\mbox{\scriptsize C\normalsize}}
\approx \frac{1}{N} \nabla_b \left( N^2
    \left(
    \begin{array}{l}
    (4C_1O^{IK} + \delta^{IK})\pi_I^a\nabla^b{A_{Ka}} + (4C_2O^{IK} -
    \delta^{IK})\pi_I^a\nabla_a{A_K^b} \\
    + 4C_3O^{IK}\pi_I^b\nabla_a{A_K^a}
    -  \left(\nabla_a{\pi_K^a}A^{Kb}
    - 2 \bar{C}^{abcd}{B^I}_{JK}   \pi_{Ia}A^J_cA^K_d)\right) \\
    - 2 O^{IK}  p_{ij} \nabla_{(d|}A_{K|l)} (2A_I^iC^{bjld} -
    A_I^bC^{ijld}) \\ - {B^I}_{JK} p_{ij} A^J_l A^K_d (
    2A_I^i\bar{C}^{bjld} - A_I^b\bar{C}^{ijld})
\end{array}
\right) \right). \label{CYMhamprop} \ee In the same sense as for
the single vector field case above, (\ref{CYMhamprop}) has the
same structure as for the GR case, so the argument used in
\cite{AB} will hold, forcing \be
    O^{IK} = \delta^{IK},
    C_1 = -C_2 = -\frac{1}{4}, C_3 = 0, \bar{C}_3 = 0, \ee \be B_{I(JK)} = 0
    \Leftrightarrow \bar{C}_1 = -\bar{C}_2 \equiv -\frac{\mbox{
    \sffamily g\normalfont}}{4}
    \label{anti}
\ee
(for some emergent coupling constant $\mbox{\sffamily g\normalfont}$) 
 and leaving the new constraint \be
    {\cal G}_J \equiv
    \nabla_a\pi^a_J - \mbox{ \sffamily g\normalfont}B_{IJK}
    \pi^I_aA^{Ka}.
\ee Again as for the single vector field case, the $\pi^a_J$
Euler--Lagrange equation is unchanged from the GR case. The action
of the dot on $A_{Ka}$ gives no volume terms.  Hence the working
for the propagation of ${\cal G}_J$ is unchanged from that in
\cite{AB}, which enforces the following conditions: \be
    I_{JLKM} = {B^I}_{JK}B_{ILM}, \bar{\bar{C}}_2 = - \bar{\bar{C}}_1
    = \frac{\mbox{ \sffamily g\normalfont}^2}{16}, \bar{\bar{C}}_3 = 0,
    \ee \be {B^I}_{JK}B_{ILM} + {B^I}_{JM}B_{IKL} + {B^I}_{JL}B_{IMK} =
    0 \mbox{ (Jacobi identity)},
\label{Jacobi} \ee \be
    B_{IJK} =
    B_{[IJK]} \mbox{ (total antisymmetry)}.
\label {totant} \ee From (\ref{anti}) and (\ref{Jacobi}), it follows
that the $B_{IJK}$ are the structure constants of some Lie
algebra, ${\cal A}$. From (\ref{totant}) and the
Gell-Mann--Glashow theorem \cite{GMG}, ${\cal A}$ is the direct
sum of compact simple and $U(1)$ subalgebras, provided that the
kinetic term is positive definite as assumed here. We can defend
this assumption because we are working on a theory in which even
the gravitational kinetic term is taken to be positive definite;
positive-definite kinetic terms ease quantization.

3) Mass terms are banned by the propagation of the Gauss laws.
Mass terms contain nontrivial powers of the volume;
however the above arguments can easily be extended to
accommodate them.  In
the many vector fields case, the effect of a mass term is to give
rise to a new term $\frac{2N}{\vc}M^{JK}A^i_K$ in the
Euler--Lagrange equations, which contributes a term
$2M^{JK}\nabla_i\left(\frac{N}{\vc}A^i_K\right)$ to the
propagation of ${\cal G}_J$.  For this to vanish, either $A^i_K =
0$ which renders the vector theory trivial, or $M^{JK} = 0$.

\section{Discussion}

The differences between the conformal-gravity--matter and
GR--matter metric Euler--Lagrange equations are the absence of a
term containing the `expansion of the universe' $p$ and the
presence of a global term such as \be -
g^{ab}\frac{\sqrt{g}}{3\vc}\left<N\left(2(R + {}^{\psi}U_{(1)}) +
3\frac{{}^{\psi}U_{(0)}}{\vc}\dots\right)\right> \ee for scalar matter.
Such a global term mimics the effect of a small epoch-dependent cosmological
constant. This global term is a `cosmological force' because it
occurs in the Euler--Lagrange equations with proportionality to
$g^{ab}$, just like the cosmological constant contribution does in
GR. We expect it to be epoch-dependent because it contains matter
field contributions, which will change as the universe evolves.
The occurrence of this global term should be compared with the
particle model in PD, in which there is a universal
cosmological force induced by all the familiar forces of nature
such as Newtonian gravity and electrostatics. We have seen that
in the particle model this cosmological
force is extremely weak over solar system scales but has a
decisive effect on cosmological scales, ensuring the conservation
of the moment of inertia, $I$.

Similarly, we expect that conformal gravity will reproduce the the solar-system 
and the binary-pulsar results\fn{We require further study of
the weak field limit to be quantitatively sure about the binary
pulsar.} just as well as GR.  This is because, first, the expansion
of the universe does not play a role on such small scales in GR so
its absence will not affect the results. Second, at maximal
expansion, a data set may be evolved by both the GR and conformal
gravity equations. The difference between these two evolutions is
well defined in Riem.  Since the first derivatives match up
at maximal expansion, the difference between the evolutions is
small.  For sure, the size of the difference will depend on the
global terms. But these can be made small by a well-known
construction, as far as the finite-time evolution for a patch of
initial data that is substantially smaller than the radius of the
universe is concerned.  Such patches can be constructed to contain
the solar system or the region containing both the solar system
and the binary pulsar.  Clearly these arguments will not apply to
cosmology, for which the differences between GR and
conformal gravity must have dramatic consequences.  

At this stage of our work, we have only just started to explore these
consequences. \it Prima facie\normalfont, it does seem
unlikely that conformal gravity will be able to supplant the
Big Bang cosmology, on account of the strong evidence
from the
Hubble redshift, nucleosynthesis and the microwave background.
Prior to further detailed study, we refer the reader to the comments
made at the end of PD. Instead, we should like
to consider the potential value of conformal gravity as a foil to
the Big Bang. Theorists concerned with achieving the deepest possible
understanding of cosmology and the foundations of physics value
alternative models \cite{conscience},
even if they explain or mimic only part of the
whole picture. For decades the Brans--Dicke alternative has played an
invaluable role, and, in its present guise as
a dilaton field, it is currently actually more orthodox than GR.
Seen in this light, conformal gravity and the reinterpretation of
the CMC-sliceable solutions of GR as geodesics on CS+V have several
positive features.

Above all, they represent a new and radical approach to scale invariance.
They show that best matching and constraint propagation are powerful
tools in theory construction. In particular, they highlight the
thought-provoking manner in which GR only just fails to be fully scale invariant.

Another potential strength of conformal gravity is that it forces one
to consider cosmology in a more sophisticated manner. Consider the
isotropic and homogeneous FRW cosmologies, which are the backbone of
the standard model. As self-similar solutions in which nothing changes
except size and homogeneous intensity, they must raise doubts. From the
dynamical point of view, they are suspiciously trivial.  In a 
scale-invariant theory, the FRW solutions 
are merely static points in the configuration space.  There has long been 
concern \cite{Krasinski} about the accuracy with which they approximate
more physically realistic inhomogeneous solutions of GR under the assumption
that it is the correct physical theory. Conformal gravity raises a more
serious doubt -- GR might be spectacularly wrong for cosmology despite
being wonderfully accurate for all other applications.  In comparison, the 
dilatonic modifications of GR \cite{Wetterich} have no significant
effect on the key physical basis of the big-bang scenario-- the explanation
of the Hubble red shift by actual expansion of the universe.

Since conformal gravity has no dynamics analogous to the FRW
universes of GR, the only possible progress in its cosmology will be
through the study of inhomogeneous solutions. This is
the opposite emphasis to the norm in classical and quantum
cosmology and does have some chance to throw up a radical new
explanation of the red shift. We know that a change in clumpiness (shape) of 
the
universe can cause redshift in GR.  The solar photons
that reach us are redshifted by having to climb out of the solar
gravitational potential well (gravitational redshift), and inhomogeneities 
cause similar
effects in cosmology (the integrated Sachs--Wolfe and Rees--Sciama
effects \cite{shapeshift}). The particle model in PD is suggestive in
this respect. We speculate that the rearrangement of
geometry and matter of an evolving universe can cause a similar redshift in
conformal gravity. In such a case, it will not be due to differences
in the gravitational potential between different points of space
but between different epochs.  Now, as pointed out in
PD, the potential can be changed either by a change of scale
or by a change of shape. Conformal gravity suggests the former is not 
available
and that the latter is the origin of the Hubble redshift. Since the
change of shape of the universe can be observed, this should lead to
testable predictions.

Another service that conformal gravity can perform is to stimulate
a thorough reexamination of the problem of singularities.
The Big Bang itself is an initial singularity where the
known laws of physics break down.  It is inevitable in GR by
theorems of Hawking \cite{Hawking}.  These require the expansion
of past-directed normal timelike geodesic congruences to be
positive everywhere on a given spatial hypersurface.  
The GR form of these theorems will not hold in conformal gravity 
since such a notion of expansion is no longer 
meaningful.\fn{ We do not know if other forms of singularity
theorem hold. We cannot so easily dismiss results involving null
and/or local expansion. Another source of trouble in adapting GR
proofs for conformal gravity will be the lack of an equivalent to the Efe's. 
  If
local singularities form, they will contribute to the global terms
in conformal gravity. This could be fatal in the particle model,
but in conformal gravity there may be two ways out. First,
singularities may only contribute a finite amount once integrated.
Second, in GR, there is the `collapse of the lapse' in approaching
local singularities, which is a gauge effect; the analogue of this
in conformal gravity would be a real physical effect. } In GR,
the Hubble redshift interpretation forces one to admit the
breakdown of known physics in our finite past, whilst in conformal
gravity, the denial of such a breakdown must be accompanied by a
new interpretation of the Hubble redshift.

Whereas our greatest interest is in whether conformal gravity can
give us an alternative cosmology, our CS+V theory has a notion
of universal expansion, so it will be much closer to GR both in agreeing
with the standard cosmology and in not offering these new 
perspectives on nonsingularity and global cosmological forces.

We finish by discussing quantization. For conformal gravity, 1) we hope
to quantize in the timeless interpretation due to one of the authors 
\cite{B94I}.
2) The Hamiltonian constraint adopts a new role in conformal gravity
since it no longer uses up a degree of freedom.  3) This and the
fundamental lapse-fixing equation (\ref{fullslicing}) are
nonstandard objects from the quantization perspective. 4) The new
global terms may play a role. 5) Finally, whereas in GR the DeWitt
supermetric gives an indefinite inner product as a consequence of
the sign of the expansion contribution to the kinetic energy, in
conformal gravity the new $W = 0$ supermetric gives a positive
definite inner product.  This ameliorates the inner product problem \cite{K93} 
of quantum gravity.  
Notice that 1) -- 3) will still be features of the quantum CS+V theory.

We have a nice set of stepping-stones toward quantization of
conformal gravity.  The effect of the $W = 0$ supermetric on
quantization can be tried out by itself in the strong gravity
regime, since one of us found this to be a consistent theory
\cite{Sanderson}.  Then the additional effect of introducing a
volume and having ${\cal H}^{\mbox{\scriptsize C\normalsize}}$ no
longer use up a degree of freedom can be tried for strong
conformal gravity, since for this the additional complication of
the lapse-fixing equation interpretation of full conformal gravity is
trivially absent (since $N$ is a spatial constant).

That conformal gravity has a
marginally smaller configuration space than GR
makes our quantum program attractive.
We hope to use a `top-down' approach:
to start from firm classical theory
and deduce features of the quantum universe.
However, we start from space rather than spacetime
for relational reasons \cite{BB, B94I, BOF, PD}
and to illustrate the potential naivety of presupposing
and extending spacetime structure.
The great problems of quantizing gravity are
hopelessly interrelated, so that adding to a partial resolution to
tackle further problems can spoil that partial resolution \cite{K93}.
So it is not to be expected that Ashtekar variable techniques
\cite{Ashtekar}, with their resolution of operator
ordering and their natural regularization, could be imported into
conformal gravity.  Thus, quantization of conformal gravity will
differ from, but not necessarily be easier than, quantization of
GR.  Should conformal gravity adequately describe the classical
universe, its quantization program will become of utmost
importance. Even if this were not the case, we expect to further
the understanding of quantization and of quantum general
relativity by such a program.

Acknowledgements:

EA is supported by PPARC and would like to thank Malcolm
MacCallum for discussions and for reading an earlier version of this script.  
N\'{O} thanks Bryan Kelleher for discussions.  EA, BF and N\'{O} thank the Barbour family for hospitality.


\begin{thebibliography}{99}
\bibitem{Einstein1}   Einstein's objection was reported by Weyl at the end of \cite{Weyltheory}.
\end{thebibliography}
\end{document}